\documentclass[review,12pt,3p]{elsarticle}
\usepackage[ruled]{algorithm2e}

\SetAlFnt{\small}
\SetAlCapFnt{\small}
\SetAlCapNameFnt{\small}
\SetAlCapHSkip{0pt}
\IncMargin{-\parindent}

\usepackage{graphicx}
\usepackage{amssymb}
\usepackage{amsmath,url}
\usepackage{multicol,multirow}
\usepackage{epsf,psfrag,pstricks}
\usepackage{rotating}
\usepackage{multirow}

\usepackage{float}
\usepackage[config,font=small]{caption}
\usepackage[scriptsize]{subfigure}
\usepackage{bm}
\usepackage{pdflscape}

\usepackage{array}
\newcolumntype{C}[1]{>{\centering\let\newline\\\arraybackslash\hspace{0pt}}m{#1}}

\journal{}
\begin{document}
\begin{frontmatter}


\title{Are You Really Hidden? Predicting Current City from Profile and Social Relationship}

\author[label1]{Xiao Han\corref{cor1}}
\ead{han.xiao@telecom-sudparis.eu}

\author[label1]{Leye Wang}
\ead{leye.wang@telecom-sudparis.eu}

\author[label2]{Jiangtao Wen}
\ead{jtwen@tsinghua.edu.cn}

\author[label1,label3]{\'Angel Cuevas}
\ead{acrumin@it.uc3m.es}

\author[label4]{Chao Chen}
\ead{cschaochen@cqu.edu.cn}

\author[label1]{Noel Crespi}
\ead{noel.crespi@telecom-sudparis.eu}

\cortext[cor1]{Corresponding author. Tel.:+33 01 60 76 41 65}
\address[label1]{Institut-Mines T\'el\'ecom, T\'el\'ecom SudParis, 9 rue Charles Fourier, 91011 Evry Cedex France}
\address[label2]{Tsinghua University, 30 Shuang Qing Lu, Haidian, Beijing, China}
\address[label3]{Universidad Carlos III de Madrid, Av de la Universidad, 30 28911 Legan\'es, Madrid, Spain}
\address[label4]{Chongqing University, 174 Shazheng Street, 400044 Shapingba, Chongqing, China}

\begin{abstract}
Privacy has become a major concern in Online Social Networks (OSNs) due to threats such as advertising spam, online stalking and identity theft. Although many users hide or do not fill out their private attributes in OSNs, prior studies point out that the hidden attributes may be inferred from some other public information. Thus, users' private information could still be at stake to be exposed. Hitherto, little work helps users to assess the exposure probability/risk that the hidden attributes can be correctly predicted, let alone provides them with pointed countermeasures. In this article, we focus our study on the exposure risk assessment by a particular privacy-sensitive attribute - \textit{current city} - in Facebook. Specifically, we first design a novel current city prediction approach that discloses users' hidden `current city' from their self-exposed information. Based on $371,913$ Facebook users' data, we verify that our proposed prediction approach can predict users' current city more accurately than state-of-the-art approaches. Furthermore, we inspect the prediction results and model the current city exposure probability via some measurable characteristics of the self-exposed information. Finally, we construct an exposure estimator to assess the current city exposure risk for individual users, given their self-exposed information. Several case studies are presented to illustrate how to use our proposed estimator for privacy protection.
\end{abstract}




\end{frontmatter}

\section{Introduction}
\label{sec:Intro} During the last decade, Online
Social Networks (OSNs) have successfully attracted billions of
people who share a huge amount of personal information through the Internet,
such as their background, preferences and social connections. Owing to the increase of potential violations such as advertising
spam, online stalking and identity theft~\cite{Gross2005}, in recent
years, more and more users have concerns about their \textit{privacy}
in OSNs and become reluctant to publish all their personal
information~\cite{FacebookTrend_PV}. Consequently, users may not fill out their privacy-sensitive attributes (e.g., location, age, or phone number), or they hide this information from strangers and only allow their friends to view such information~\cite{Chen2013661}.

While hiding the privacy-sensitive attributes, users usually expose
some other information that appears to be less sensitive to them. It
has been reported that Facebook users publicly reveal four
attributes on average, and $63\%$ of them uncover their friends list
\cite{analysis_public_info}. Due to the correlations among various
attributes, some of the self-exposed information may indicate the
invisible privacy-sensitive attributes to some extent~\cite{CCP}\cite{ICDMW2012}. Hence, it is questionable whether the privacy-sensitive attributes that a user intends to hide are really hidden.

This work, using location information as a representative case, aims to assess what is the risk that a user's invisible information could be disclosed. There are several reasons that lead us to conduct this study based on location information. First, among various kinds of information, location is
usually one of the privacy-sensitive attributes for most users~\cite{chakraborty2013privacy}. In
real-life OSNs, we notice that users are quite careful to not reveal
their location information: $16\%$ of users in Twitter reveal home
city \cite{MLP} and $0.6\%$ of Facebook users publish home
address~\cite{find_me}. Second, location
information is a commercially valuable attribute which might even be misused by unscrupulous businesses to bombard a user with unsolicited marketing~\cite{duckham2006location}.
In addition, location information leakage may lead to a spectrum of intrusive inferences such as inferring a user's political view or personal preference~\cite{duckham2006location}\cite{han2015alike}.
Therefore, protecting the hidden location information for a user
becomes rather critical. In particular, as Facebook is the most
popular OSN~\cite{SM_report}, we concentrate on the attribute of
\textit{current city} in Facebook and investigate the following
issues:

\vspace{0.1cm} 1) \textit{Is the private current city that a user
expects to hide really hidden? In other words, if a user hides
his current city but exposes some other information, can we predict
a user's current city by using his self-exposed information?}

\vspace{0.15cm} 2) \textit{Can we help individual users
to understand the actual risk (probability) that their private current
city could be correctly predicted based on their self-exposed
information? Furthermore, can we provide some countermeasures to
increase the security of the hidden current city?} \vspace{0.1cm}

To address these issues, we first propose a current
city prediction approach to predict users' hidden current city.
Although many location prediction approaches have been developed for
Twitter~\cite{T_content_prediction}\cite{geo_twitter}\cite{insights}\cite{www2014}
and Foursquare~ \cite{ICDMW2012}\cite{foursquare_behavior}, they
cannot be appropriately implemented on Facebook because of the
different properties (e.g., obtainable information) in these OSNs.
For Facebook, Backstrom et al. predict users' locations based on
their friends' locations~\cite{find_me}. In addition to friends' locations,
users' profile attributes, such as hometown, school and workplace,
may also indicate their current city to some extent~\cite{CCP}. In
order to achieve high prediction accuracy in Facebook, we devise a
novel current city prediction approach by extracting location
indications from integrated self-exposed information including
profile attributes and friends list.

Second, based on the proposed prediction approach, we construct a
current city exposure estimator to estimate the exposure probability
that a user's invisible current city may be correctly inferred via
his self-exposed information. The exposure estimator can also
provide a user with some countermeasures to keep his hidden current
city hidden. To the best of our knowledge, this is the first
work that estimates the exposure probability of a user's invisible attribute by his self-exposed information.


It is a non-trivial task to construct either the current city
prediction approach or the exposure estimator. We encounter the
following challenges:

1) \textbf{\textit{How to extract and integrate different location
indications from a user's multiple self-exposed information?}} Since
the proposed prediction approach explores location
indications from both profile attributes and friends list, two
subproblems are considered. $(i)$~A user probably reveals multiple
attributes (e.g., hometown, workplace) which may indicate different
locations; besides, a certain attribute might indicate several
locations. For example, a user working in \textsc{Google} suggests
that the user could probably live in any city where \textsc{Google}
sets up an office, e.g., \textsc{California}, \textsc{Beijing} or
\textsc{Paris}. $(ii)$~The friends of a user, probably residing in
different cities, may be close to or far away from the user. These
strong or weak geographic relations may influence the significance
of the friends' location indications. Thus, it is challenging to
appropriately combine these various location indications into an
integrated model, so as to determine the probabilities of locations
where the user may live.\color{black}


2) \textbf{\textit{How to predict a user's current city when we
obtain the probabilities of the user being at various locations?}}
By overcoming \textit{challenge 1}, we can obtain a probability
vector which indicates the probabilities that a user resides at
certain locations. With this probability vector, a straight-forward prediction approach could select the location with the highest probability as the user's current city. However, this might not be the best option
when concerning the locations' geographic relations. Assume the
probability vector suggests that a user $u$ has $40\%$, $35\%$ and
$25\%$ probability of residing in \textsc{Beijing}, \textsc{Paris}
and \textsc{Evry} respectively. Then, $u$ is more likely to live in
the area around \textsc{Paris} and \textsc{Evry} than
\textsc{Beijing}, because \textsc{Paris} and \textsc{Evry} are only
$30km$ apart but they are thousands of kilometers away from
\textsc{Beijing}. Hence, a location selection method should be
carefully designed for a current city prediction approach.

3) \textbf{\textit{How to estimate the exposure risk of a user's
hidden current city?}} To help a user understand the exposure risk
of his hidden current city, a straight-forward method would be to provide a predicted location; thus the user can decide whether
his current city can be predicted correctly (risky) or incorrectly
(secure). However, this method may not meet users' expectations. A user, whose location is correctly predicted, may expect being able to know which of his self-exposed information primarily leads to the leakage of his private current city and how to increase its security. A user, whose hidden location is not predicted correctly, still needs to be aware of some leakage of location that may exist. For example, a
prediction approach may incorrectly infer a Parisian living in
\textsc{Lyon} according to probabilistic results: $55\%$ in
\textsc{Lyon} and $45\%$ in \textsc{Paris}; Even though the
prediction result is incorrect, the user still leaks some location
information. Therefore, how to estimate the current city exposure
risk and help a user achieve his desired privacy level is a challenging
objective.

This paper makes the following contributions:

1) \textbf{Profile and friend location indication model:} To
properly extract location indications from users' self-exposed
information, we construct an integrated probability model. We
capture location indications from two types of information:
\textit{location sensitive attributes} and \textit{friends list}.
Location sensitive attributes are the profile attributes that can
indicate one or multiple locations. In this paper, we use `Hometown'
and `Work and Education' as the location sensitive attributes. For
each location sensitive attribute, we set up a \textit{location
attribute indication matrix} from which we can index the locations
and the corresponding probabilities that a certain attribute value
indicates. Besides, considering a user and each of his friends who
publish current city, we estimate their location similarity
according to their attribute correlations, and assign a large weight
to a friend that has a high location similarity to the user. For
a friend who does not reveal current city, we predict the friend's current
city using his visible location sensitive attributes, and assign him a
very small weight. Finally, based on information from $371,913$ users collected from Facebook, we train an integrated model that can determine the probability for each potential city where a user may
reside.

2) \textbf{Current city prediction approach:} To address
\textit{Challenge 2}, we aggregate locations into clusters by
considering the locations' geographic relations. Then, based on the
proposed \textit{profile and friend location indication model}, we
predict a user's invisible current city in two steps: $(i)$
\textit{cluster-selection}: for each cluster, we sum up the probabilities of
locations inside the cluster; then we select the cluster with the
highest probability; $(ii)$ \textit{location-selection}: we determine a best
location within the selected cluster as the user's current city. The
evaluation results demonstrate that our proposed prediction approach
achieves lower error distance and higher accuracy than the
state-of-the-art approaches. Furthermore, for the users who reveal
their `Hometown' and `Work and Education', our proposed approach can
predict current city with an accuracy of $90\%$.

3) \textbf{Current city exposure estimator:} We define some
measurements to describe the characteristics of users' self-exposed
information.
Based on these measurements, we analyze how the users' self-exposed
information affects the probability that their current city may be
correctly inferred (i.e., current city exposure probability).
Furthermore, \textit{Random Decision Forest} method is employed to model the current city
exposure probability, and subsequently a current city exposure
estimator is constructed. Given a user's self-exposed information, the proposed
exposure estimator provides two estimators
--- \textit{Exposure Probability} and \textit{Risk Level} --- to
quantify the current city exposure risk. The exposure estimator can
also estimate the exposure risk assuming that the user hides some of
his self-exposed information. Consequently, the user can easily
decide which information he should hide to satisfy his privacy
intention.

The rest of this paper is organized as follows. We review the
literature in Sec.~\ref{sec:Literature}, formulate the current city
prediction problem in Sec.~\ref{sec:problem}, and overview our
solution to the prediction problem in Sec.~\ref{sec:overview}. Next, the profile and friend location indication model is devised in Sec.~\ref{sec:model}; the current city prediction approach is respectively presented and evaluated in Sec.~\ref{sec:approach} and Sec.~\ref{sec:Eva}. By inspecting the current city prediction results, Sec.~\ref{sec:exposure} proposes the exposure estimator. Finally, Sec.\ref{sec:Discussion} makes some discussions and points out future work. Sec.~\ref{sec:conclusion} concludes this work.



\section{Literature Review}
\label{sec:Literature} In this section, we briefly review the
related work from two perspectives: city-level
location prediction and privacy in OSNs.
\subsection{City-Level Location Prediction}
Existing city-level location prediction approaches can be classified into
four categories: \textit{relationship-based} prediction,
\textit{content-based} prediction, \textit{hybrid content-relationship} prediction and \textit{multi-indication} prediction.

\subsubsection{Relationship-based Prediction}
Based on the principle that the probability of being friends is
declining with geographic distance, this prediction category infers
a user's location according to the visible locations of his friends~\cite{find_me}. Researchers have studied the correlation between
geographic distance and social relationship on large-scale Facebook
users in United States. They reveal that the probability of being
friends falls down monotonically as the distance increases. Depending
on this observation, they build a maximum-likelihood location
prediction model and finally refine the prediction with an
iterative algorithm.

\subsubsection{Content-based Prediction}
The rise of Twitter has spawned a mass of tweets. As some tweets
contain location-specific data, this category of prediction
approaches~\cite{T_content_prediction}\cite{geo_twitter}\cite{insights}
infers a user's location relying on his location-related tweets. The
basic idea of these approaches is to detect the location-related
tweets and construct a probabilistic model to estimate the
distribution of location-related words used in tweets. In order to
raise the prediction accuracy, the basic idea is improved by various means, such as such as selecting the top K probable
cities~\cite{T_content_prediction}, identifying words with a strong
local geo-scope and refining the prediction with a neighborhood
smoothing model~\cite{geo_twitter}.

\subsubsection{Hybrid Content-Relationship Prediction}
Another compelling category combines the location indications from
relationships and tweet content. TweetHood identifies a user's
location by exploring both his tweets and his closest friends'
locations~\cite{TweetHood}. Tweecalization improves TweetHood by
employing a semi-supervised learning algorithm and introducing a new
measurement which combines trustworthiness and the number of common
friends to weight friends~\cite{Tweecalization}. Li et al. integrate
the location influences captured from both social network and
user-centric tweets into a unified discriminative probabilistic
model~\cite{UID}. By considering a user who may be related to multiple
locations, MLP model~\cite{MLP} proposes to set up a complete
`location profiles' prediction which infers not only a user's home
location but also his other related locations.


\subsubsection{Multi-Indication Prediction}
Besides users' relationships and content, multi-indication
prediction approaches explore multiple location indications from
other possible location resources to infer users' invisible
location. To resolve ambiguous toponymies in tweet content, besides
location indications extracted from tweets, existing work has
introduced location indications from websites' country code,
geocoded IP addresses, time zone and UTC24-offset
\cite{Multi_Indicator}. Such a multi-indication idea has also been
used to Foursquare, which specifically exploits mayorships, tips and
dones that users marked \cite{foursquare_behavior}. However, all
these multi-indication prediction approaches are proposed for either Twitter
or Foursquare, but not for Facebook. Our previous work reveals the
statistical analyzed correlation between users' current city and
other location sensitive attributes in Facebook~\cite{CCP}. It also
predicts a user's current city with city-level and country-level
results by using a neural network approach. However, this previous
work assumed that an attribute value could map to a specific
location, which is not true for many cases. Recall the example that
a user works in \textsc{Google} might work in \textsc{California},
\textsc{Paris} or \textsc{Beijing} (\textit{Challenge 1} of Sec.~\ref{sec:Intro}).

In this paper, we consider multiple location indications by
integrating relationship and profile attributes in Facebook.
Compared to our previous work where an attribute value only allows
to bind with one fixed location~\cite{CCP}, an attribute value can
be mapped to multiple locations with different probabilities in the
newly proposed model. In addition, we consider both the friends
whose current city is either visible or invisible; whereas the
existing work relies on the friends who reveal their
locations~\cite{find_me}\cite{UID}. Particularly, we propose a new
approach to bias the weights of friends whose current city is
visible.

\subsection{Privacy in OSNs}
In OSNs, users are more and more concerned with privacy of their
personal information~\cite{FacebookTrend_PV}. A majority of users
configure their privacy settings and hide some of their information
from strangers. Unfortunately, previous research has pointed out the
disparity between the expectation and the reality of users' privacy;
and it has also showed that much of users' private information is easily uncovered~\cite{fb_privacy}.

Much existing work ascribes the privacy leakage to the users themselves.
On one hand, users might incorrectly manage their privacy settings
due to the poor human-computer interaction or complex privacy
maintainability~\cite{fb_privacy}\cite{chakraborty2013privacy}. To address this issue, researchers have
designed a user-friendly interface for managing privacy settings
with an audience view~\cite{design_interface}.

On the other hand, users only hide some of the attributes that are
privacy-sensitive to them while make the others accessible to
public --- users on Facebook generally expose more than four attributes to
strangers and $63\%$ of users share their friend lists with the
public \cite{analysis_public_info}. As reported, such user
self-exposure behavior leaves a huge chance for inferring the
hidden attributes
\cite{age_estimate}\cite{infer_prof}\cite{Infer_PV}. Many tools have
been developed to infer users' invisible information by various
means such as inferring the private information through users' other
self-exposed information \cite{infer_prof}, their social connections
\cite{age_estimate}\cite{Infer_PV} and social
groups~\cite{Illusion_PV}\cite{User_Prof}.

Some papers claim that it is hard for a user to avoid privacy
leakages if he only hides the private attribute
~\cite{age_estimate}\cite{infer_prof}\cite{Strategies_struggles};
whereas many studies merely suggest users with a general idea of
hiding other attributes so as to become more secure (e.g., hide
relationships~\cite{protect_privacy}). Unlike the above
work, we provide an individual user with the exposure probability of
his private current city concerning his self-exposed information.
We also suggest some pointed rules for protecting users'
privacy on their current city.

\section{Formulation of Current City Prediction Problem}
\label{sec:problem} In this section, we formulate the current city
prediction problem. Facebook, as a social network containing
location information, can be viewed as an undirected graph
$\mathcal{G}=(\mathcal{U},\mathcal{E},\mathcal{L})$, where
$\mathcal{U}$ is a set of users; $\mathcal{E}$ is a set of edges
$e\langle u, v \rangle$ representing the friend relationship between
users $u$ and $v$, where $u$ and $v \in \mathcal{U}$; $\mathcal{L}$
is a candidate locations list composed of all the user-generated
locations.

Typically, a user $u$ in Facebook might contribute various items of information, e.g., basic profile information, friends, comments and photos. The core information of $u$ in this paper is the user's current
city, denoted as $l(u)$. The users are classified into two sets according to the accessibility of users'
current city: current city
available users (LA-users) and current city unavailable users
(LN-users). We, respectively, use $\mathcal{U}^{^{LA}}$ and
$\mathcal{U}^{^{LN}}$ to denote the sets of LA-users and LN-users,
where $\mathcal{U}=\mathcal{U}^{^{LA}} \cup \mathcal{U}^{^{LN}}$.

To predict users' current city, we exploit the users'
location sensitive attributes and friends list. Assume that there
exist $m$ types of location sensitive attributes, denoted as
$\mathcal{A} = \{a_1, a_2, \cdots, a_m\}$. Specifically, we denote a
user $u$'s location sensitive attributes as $\mathcal{A}(u) =
\{a_1(u), a_2(u), \cdots, a_m(u)\}$. The users may also have a
friends list, denoted as $\mathcal{F}(u)$, where $\mathcal{F}(u) =
\{f \in \mathcal{U}: e \langle u, f \rangle \in
\mathcal{E}\}$. Therefore, we use a tuple to represent a user as $u:
\langle l(u), \mathcal{A}(u), \mathcal{F}(u) \rangle$.

Additionally, each location is associated with a unified \textit{ID} ($l_{id}$). Then,
with this \textit{ID}, we can obtain each location's latitude and longitude
coordinate via Facebook Graph API Explorer. Therefore, a
location can also be written as a tuple: $l : \langle l_{id}, lat, lon
\rangle$ and the candidate locations list can be denoted as a set of
location tuples: $\mathcal{L} = \{l : \langle l_{id}, lat, lon
\rangle \}_N$, where $lat$ and $lon$ respectively stand for the
latitude and longitude of a location, and $N$ is the number of
candidate locations in the list.

Thus, the \textbf{\textit{current city prediction problem}} can be formally
stated as: \textit{Given, (i)~a graph $\mathcal{G}=(\mathcal{U}^{^{LA}}\cup \mathcal{U}^{^{LN}},\mathcal{E},\mathcal{L})$; (ii)~the public location $l(u)$ for LA-users $u \in U^{^{LA}}$; (iii)~the location sensitive attributes $\mathcal{A}(u)$ and the friends list $\mathcal{F}(u)$ for all the users $u \in (\mathcal{U}^{^{LA}} \cup \mathcal{U}^{^{LN}})$, we predict current city $\hat{l}(u)$ for each LN-user $u \in \mathcal{U}^{^{LN}}$, so as to make $\hat{l}(u)$ close to the user's real current city.}

Note that the current city of a user's friends can be either
available ($f \in \mathcal{U}^{^{LA}}$) or unavailable ($f \in \mathcal{U}^{^{LN}}$).
Thus, we introduce two notations to represent the two groups of
friends: current city available friends (LA-friends) and current
city unavailable friends (LN-friends). Let denote a user's
LA-friends as $\mathcal{F}^{^{LA}}(u)$ and LN-friends as
$\mathcal{F}^{^{LN}}(u)$, where $\mathcal{F}(u)
=\mathcal{F}^{^{LA}}(u) \cup \mathcal{F}^{^{LN}}(u)$.

\section{Overview of Current City Prediction}
\label{sec:overview} 
\begin{figure}[!htp]
\centering
\includegraphics[width=9cm, height=5.8cm]{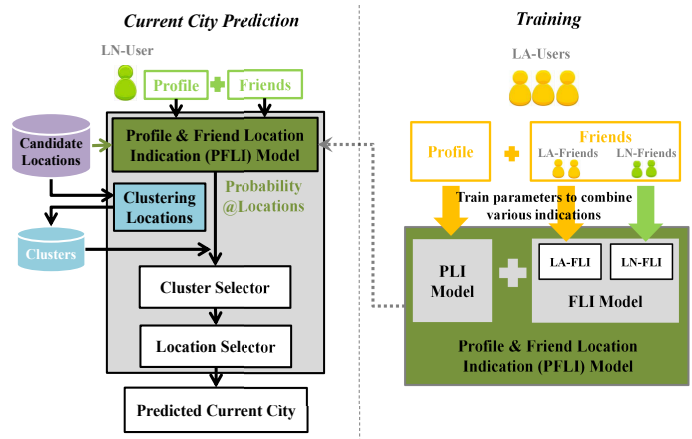}
\caption{Framework of Current City Prediction.}
\label{fig:overview}
\vspace{-0.5cm}
\end{figure}

The goal of current city prediction is to correctly infer a coordinate point with latitude and longitude for a LN-user, given the candidate locations list $\mathcal{L}$ and the
user's self-exposed information including his location sensitive attributes and friends list. 
Figure~\ref{fig:overview} illustrates the framework of the proposed current city prediction solution. To determine the current city of a LN-user, we first train an integrated profile and friend location indication (i.e., \textit{PFLI}) model to compute the probabilities of the candidate locations in which the LN-user may currently live. Next we take a two-step location selection strategy: cluster selection and location selection. Specifically, we aggregate the nearby locations into a location cluster and obtain a set of location clusters. We then calculate the probability of a user being in a cluster by summing up the probabilities of all the candidate locations belonging to this cluster; the cluster with the
highest probability is picked out as a candidate cluster. Finally, we try to select the `best' location from the candidate cluster as the predicted current city.

To train the integrated \textit{PFLI} model (see the right-hand part of Figure~\ref{fig:overview}), we separately consider the location indications from location sensitive attributes and friends, and consequently obtain two sub-models: profile location indication (\textit{PLI}) model and friend location indication (\textit{FLI}) model. Both \textit{PLI} model and \textit{FLI} model calculate a probability vector in which the element stands for the probability of a user being at a certain candidate location. Note that, \textit{FLI} model leverages the location indications from both LA-friends and LN-friends. By integrating the probability vectors that are generated by \textit{PLI} and \textit{FLI} models with appropriate parameters, a unified profile and friend location indication (\textit{PFLI}) model is derived.

Next, we will elaborate the \textit{PFLI} model and the current city prediction approach.

\section{Profile and Friend Location Indication Model}
\label{sec:model} In this section, we describe the design of the
probabilistic models that can suggest the probabilities of users
being at each of the candidate locations. We first introduce the
\textit{profile location indication} (\textit{PLI}) model; it
estimates the probability of each candidate location by merely relying on a user's location
sensitive attributes. Then, we describe the \textit{friend location
indication} (\textit{FLI}) model, which captures the location
indications from a user's friends. Finally, we integrate these
two models and obtain the integrated \textit{profile and
friend location indication} (\textit{PFLI}) model.

\subsection{Profile Location Indication Model}
\label{sec:PLI} According to \textit{Challenge 1} in Sec.
\ref{sec:Intro}, two problems should be considered in constructing
\textit{PLI} model. First, a certain value of a location sensitive
attribute may indicate several locations. For instance, \textsc{Google}, being a certain value of workplace, could indicate any city where \textsc{Google} sets up an office such as \textsc{California}, \textsc{Beijing} or \textsc{Paris}. Therefore, for each attribute value, we consider all possible location
indications with the corresponding probabilities. Second, a user may present multiple
location sensitive attributes (e.g., hometown, workplace, college). Thus we integrate
various location indications extracted from different location
sensitive attributes.

To capture the multiple possible location indications from
one attribute value, we define a \textit{location-attribute
indication matrix} for each ($k$-th) location sensitive attribute
$a_k \in \mathcal{A}$, denoted as $\mathcal{R}_{k}$. The rows of
this matrix represent the candidate locations ($l \in \mathcal{L}$),
while the columns stand for the possible values of $a_k$. We use
$l_i$ to represent the $i$-th candidate location and $a_{k_j}$ to
denote the $j$-th possible value of $a_k$. A cell
$\sigma_{k}^{ij}$ in the matrix calculates the \textit{indication
probability} of $a_{k_j}$ to $l_i$ --- the probability that a user,
whose $k$-th location sensitive attribute $a_k$ equals $a_{k_j}$,
currently lives in the city $l_i$. Specifically, the indication
probability equals the number of users who live in $l_i$ and have a
value of $a_{k_j}$ divided by the total number of users who have a
value of $a_{k_j}$. For instance, considering workplace, if $10$ out
of $100$ employees from \textsc{Telecom SudParis} in the whole data set state that they
live in \textsc{Evry}, then the indication
probability of \textsc{Telecom SudParis} to \textsc{Evry} is $0.1$.
Note that, the $j$-th column of $\mathcal{R}_{k}$ represents the
multiple location indications of $a_{k_j}$.

Assume that $a_k$ has $M$ possible values except \textit{null}; $N$
is the total number of the candidate locations. The $k$-th
location-attribute indication matrix can be written as:
\begin{align*}
\mathcal{R}_{k}=\{\sigma_{k}^{ij}\}_{N \times
M}= \{p(l(u)=l_i | a_k(u)=a_{k_j})\}_{N
\times M}=[R_{\cdot
k_1}, R_{\cdot k_2},...,R_{\cdot k_M}]
\end{align*}
where $R_{\cdot k_j}$ represents all the locations' probabilities for a user who presents $a_{k_j}$.

Based on the location-attribute indication matrix ($\mathcal{R}$),
we model the probability of a user's current city at $l_i$ by
combining all of a user's available location sensitive attributes in
his profile:
\begin{equation}
\begin{split}
\label{eq:PLM1}
p_{_{Prof}}(u,l_i) &= \sum_{a_k \in \mathcal{A}, a_{k}(u) \neq \textit{null}}{\alpha_k p(l(u)=l_i | a_{k}(u)=a_{k_j})} \\
&= \sum_{a_k \in \mathcal{A},a_{k}(u) \neq \textit{null}}{\alpha_k
\sigma_{k}(u,l_i)}
\end{split}
\end{equation}
where $\sigma_{k}(u,l_i)$ can be easily obtained by indexing the
corresponding location-attribute indication matrix
($\mathcal{R}_{k}$) according to $u$'s value of $a_k$
($a_k(u)=a_{k_j}$) and the given location ($l_i$), namely
$\sigma_{k}^{ij}$; $\alpha_k$ is a parameter to adjust the
significance of the different location sensitive attributes.

As we discussed in Sec. \ref{sec:problem}, a user may not reveal some attributes. Therefore, in Eq.
\ref{eq:PLM1}, the location indication from the attribute $a_k(u)$ at any location equals zero if the user's $a_k(u)$ is invisible. If all of a user's
location sensitive attributes are invisible, we rely on
his friends' information to infer his current city,
which we will discuss in the next section.

\subsection{Friend Location Indication Model}
In addition to a user's location sensitive attributes, 
we explore location indications from users' friends to construct \textit{FLI} model. A user's friends can be either LA-friends (current city available) or LN-friends (current city unavailable). We build up \textit{FLI} model primarily depending on LA-friends' location indications and also considering LN-friends' location indications as a small regulator. Accordingly, \textit{FLI} model contains two components: LA-friends location indication (\textit{LA-FLI}) model and LN-friends location indication (\textit{LN-FLI}) model.

\subsubsection{LA-FLI Model} \textit{LA-FLI} model differentiates the weights of a user's LA-friends and estimates his probability of living in location $l_i$ by the weights of his friends who also live in $l_i$. \textit{LA-FLI} model expects to assign high weights to the LA-friends who live in the same city as the user does. However, since the user's city is unknown, whether or not a friend and the user live in the same city cannot be directly determined. Therefore, \textit{LA-FLI} model assesses the likelihood that two users live in a same city (i.e., \textit{location similarity}) based on the correlation between their location sensitive attributes.
\begin{figure}[!htp]
\centering
\includegraphics[width=6.5cm, height=3.5cm]{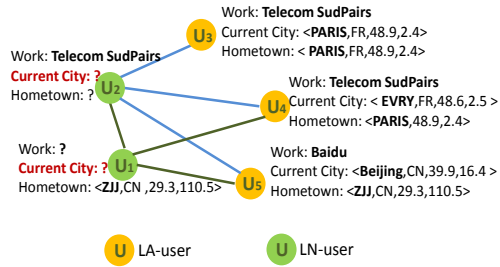}
\caption{An Example of Social Relations and Profile Information.}
\label{fig:example}
\end{figure}
Figure \ref{fig:example} illustrates an example to show that
the location sensitive attributes can be used to distinguish the
weights among various LA-friends. Focusing on LN-user $u_2$ and his
LA-friends $u_3$, $u_4$ and $u_5$, we notice that $u_2$ and $u_3$,
$u_4$ work in the same institute, while $u_5$ works in another
company which is far away from $u_2$'s workplace. In this case, it
is natural to infer that $u_2$ is more likely to be living in the
same city with $u_3$ and $u_4$ than with $u_5$; then $u_3$ and $u_4$
should be assigned with higher weights than $u_5$ because of the
location similarity indicated by their workplace.


Inspired by the example, we construct an \textit{attribute-based
location similarity matrix} ($\mathcal{W}_{k}$) by each ($k$-th) location
sensitive attribute ($a_k \in \mathcal{A}$). In the matrix, a cell $w_{k}^{ij}$ calculates the probability that two users live in the same city (i.e., \textit{location similarity}) when they respectively have values of $a_{k_i}$ and $a_{k_j}$ regarding $a_k$.
Specifically, we compute the total number of friend pairs
where one user has a value of $a_{k_i}$ and the other has a value of
$a_{k_j}$, denoted as $|\{a_k(u)=a_{k_i} \wedge a_k(v)=a_{k_j}\}|$;
Among these friend pairs, we further count the pairs of friends who live in the same city, denoted as $|\{l(u)=l(v) \wedge a_k(u)=a_{k_i} \wedge a_k(v)=a_{k_j}\}|$. Then,
\begin{align*}
\mathcal{W}_{k} &= \{w_{k}^{ij}\}_{M \times M}\\
&= \{p(l(u)=l(v) | a_k(u)=a_{k_i} \wedge
a_k(v)=a_{k_j})\}_{M \times M}\\
&= \{\frac{|\{l(u)=l(v) \wedge a_k(u)=a_{k_i} \wedge a_k(v)=a_{k_j}\}|}{|\{a_k(u)=a_{k_i} \wedge a_k(v)=a_{k_j}\}|}\}_{M \times M}
\end{align*}
where $M$ is the number of possible values of attribute $a_k$ including $null$.

For a certain attribute $a_k$, assume that $u$ and his LA-friend $v$
have a value of $a_{k_i}$ and $a_{k_j}$ respectively. Then, the
$u$ and $v$'s location similarity on $a_k$ can be
easily obtained by indexing the $i$-th row and $j$-th column of
$\mathcal{W}_{k}$, denoted as $w_{k}(u,v) = w_{k}^{ij}, v \in
\mathcal{F}^{^{LA}}(u)$.

We combine multiple location similarities on all the location sensitive
attributes (e.g., work, hometown) with a set of trained parameters ($\mathbf{\beta}$) to
measure $v$'s weight. This combined weight describes the probability that $u$ and $v$ live in the same city concerning all of their location sensitive
attributes.

Then, \textit{LA-FLI} model calculates the probability of $u$ living
in $l_i$ by integrating all the weights of $u$'s LA-friends who live
in $l_i$:
\begin{equation}
\label{eq:LA-FLI} p_{_{LA-F}}(u,l_i)=\sum_{v \in
\mathcal{F}^{^{LA}}(u)}\sum_{a_k \in
\mathcal{A}}{\beta_k{w_{k}(u,v)p_{_{LA-U}}(v,l_i)}}
\end{equation}
where $p_{_{LA-U}}(v,l_i)$ represents whether or not the LA-friend
$v$ living in $l_i$. It equals $1$ if $v$ states his current city is
$l_i$; otherwise, it is $0$:
$$
p_{_{LA-U}}(v,l_i) =
  \begin{cases}
   1 & \textit{if } l(v) = l_i \\
   0 & \textit{otherwise}
  \end{cases}
$$

\subsubsection{LN-FLI Model}
Before introducing \textit{LN-FLI} model, we inspect the potential benefit of
a user's LN-friends for his current city prediction with another
example shown in Figure~\ref{fig:example}. We observe that $u_2$,
being a LN-friend of $u_1$, does not expose his current city;
whereas, the workplace of $u_2$, \textsc{Telecom SudParis},
indicates two cities
--- \textsc{Paris} and \textsc{Evry}
--- according to the current cities of the users $u_3$ and $u_4$ who
are also the employees of \textsc{Telecom SudParis}. Thereby, a
user's LN-friends can also reveal some location indications in their
exposed attributes, which may help the prediction.

Therefore, for a LN-friend $v$, we first rely on his exposed
location sensitive attributes and use \textit{PLI} model (Sec.
\ref{sec:PLI}) to predict his current city, as:
$$
p_{_{Prof}}(v,l_i) = \sum_{a_k \in \mathcal{A},a_k(v) \neq
\textit{null}}{\alpha_k}p(l(v)=l_i | a_k(v)=a_{k_j})
$$

Treating all the LN-friends equally, \textit{LN-FLI} model integrates
LN-friends' location indications and computes the probability that
$u$ lives in $l_i \in \mathcal{L}$ as:
\begin{equation}
\label{eq:LN-FLI} p_{_{LN-F}}(u,l_i)=\sum_{v \in
F^{LN}(u)}p_{_{Prof}}(v,l_i)
\end{equation}

\subsubsection{FLI Model}
Finally, primarily relying on \textit{LA-FLI} model and being
adjusted by \textit{LN-FLI} model with a small regulator
parameter $\lambda$, \textit{FLI} model estimates the probability
that $u$ currently lives in $l_i$ as:
\begin{equation}
\label{eq:FLI} p_{_F}(u,l_i)=p_{_{LA-F}}(u,l_i)+\lambda
p_{_{LN-F}}(u,l_i)
\end{equation}

\subsection{Integrated Profile and Friend Location Indication Model}

Next, we discuss how to integrate \textit{PLI} model and \textit{FLI}
model into a unified probabilistic location indication model, so as to capture the complete
location indications. Specifically, \textit{PFLI}
model calculates the probability of $u$ living in $l_i \in
\mathcal{L}$ as:
\begin{equation}
\label{eq:PFLI} p(u,l_i)=\theta_{_{P}}p_{_{Prof}}(u,l_i)+\theta_{_F} p_{_F}(u,l_i)
\end{equation}

\textbf{Parameter Computation:} To obtain a set of good parameters
for the model, we first rewrite the model as:
\begin{equation}
\begin{split}
\label{eq:FLI_cmpt}
 p(u,l_i)&= \sum_{a_k \in \mathcal{A}}{\theta_{_{P}} \alpha_k
\sigma_{k}(u,l_i)}\\
&+\sum_{a_k \in \mathcal{A}}{\theta_{_{F}} \beta_k \sum_{v \in \mathcal{F}^{^{LA}}(u)} {w_{k}(u,v) p_{_{LA-F}}(v,l_i)}}\\
&+ \sum_{a_k \in \mathcal{A}} \lambda \theta_{_{F}} \sum_{v \in \mathcal{F}^{^{LN}}(u)} \alpha_k \sigma_{k}(v,l_i)\\
&=\sum_{a_k \in \mathcal{A}} \{[\mu_{k} \sigma_{k}(u,l_i)+\nu_{k}
\delta_{k}(u,l_i)] + [\lambda_{\alpha} \alpha_k \eta_{k}(u,l_i)]\}
\end{split}
\end{equation}
where
\begin{itemize}
\item $\mu_{k}=\theta_{_P} \alpha_k$; $\nu_{k}=\theta_{_F} \beta_k$; $\lambda_{\alpha}=\lambda \theta_{_F}$
\vspace{+.1cm}
\item $\delta_{k}(u,l_i) = \sum_{v \in \mathcal{F}^{^{LA}}(u)} {w_{k}(u,v) p_{_{LA-F}}(v,l_i)}$
\vspace{+.1cm}
\item $\eta_{k}(u,l_i) = \sum_{v \in \mathcal{F}^{^{LN}}(u)} \sigma_{k}(v,l_i)$
\end{itemize}

The location indications extracted from a user's location sensitive
attributes and his LA-friends are considered as primary indications, while the location indication
captured from the LN-friends is only used to regulate the results. Therefore, we integrally train a good set of parameters $\mu_k$ and $\nu_{k}$; while we separately train $\alpha_{k}$.

To train the parameters $\mu_k$ and $\nu_{k}$, we generate a training data set with items $\langle \text{label}(l_i): \text{features}(u,l_i) \rangle$, if the probability that a LA-user $u$ lives in
$l_i$ is larger than zero, i.e., $\sum_{a_k \in \mathcal{A}}{[\sigma_k(u,l_i) + \delta_k(u,l_i)] > 0}$. In particular, $l_i$ is labeled as a \textsl{far} location ($\text{label}(l_i)=0$), if the distance between $l_i$ and $u$'s actual location is larger than a pre-defined threshold; otherwise, it is labeled as a \textsl{close} location ($\text{label}(l_i)=1$). Additionally, $\text{features}(u,l_i)$ is a vector consisting of $\sigma_{k}(u,l_i)$ and $\delta_{k}(u,l_i)$, where $k \in [1,m]$ represents the $k$-th location sensitive attribute. Based on the generated items, we use a logistic regression
method to train the model in the following format:
$$f(y|\mathbf{x};\sigma_1,\cdots,\sigma_m,\delta_1,\cdots,\delta_m)=h_{\mathbf{\sigma},\mathbf{\delta}}(\mathbf{x})^y(1-h_{\mathbf{\sigma},\mathbf{\delta}}(\mathbf{x}))^{1-y}$$
where $y$ is the $\text{label}(l_i)$, $\mathbf{x}$
stands for the $\text{features}(u,l_i)$ and
$h_{\mathbf{\sigma},\mathbf{\delta}}(\mathbf{x})$ is the hypothesis
function. Then we can apply the gradient descent method to maximize
$f(y|\mathbf{x};\mathbf{\sigma},\mathbf{\delta})$ and compute the
parameters. In the similar way, we can train a set of parameters $\alpha_{k}$.

\section{Current City Prediction Approach}
\label{sec:approach} To address \textit{Challenge 2} of Sec. \ref{sec:Intro}, we aggregate the close candidate locations into clusters and devise a two-step current city selection approach. In this section, referring to Figure \ref{fig:overview}, we elaborate the \textit{Candidate Locations Cluster}, \textit{Cluster Selector} and \textit{Location Selector} respectively. We summarize the prediction approach at the end of this section.

\subsection{Candidate Locations Cluster}
\label{sec:lcc}

We draw on the hierarchical clustering method, i.e., UPGMA (Unweighted Pair Group Method with Arithmetic Mean)~\cite{sokal1958statistical}\cite{hastie01statisticallearning}, to generate location clusters. This method arranges all the candidate locations in a hierarchy with a treelike structure based on the distance between two locations, and successively merges the closest locations into clusters. Algorithm \ref{alg:cluster} elaborates the clustering process.\color{black}

\begin{figure}[!htp]
\centering
\includegraphics[width=8.5cm, height=5.5cm]{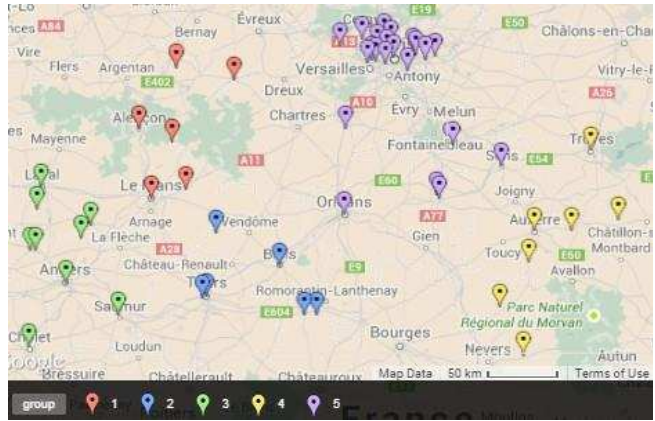}
\caption{Example of Candidate Locations Cluster.}
\label{fig:cluster}
\end{figure}

Figure \ref{fig:cluster} illustrates an example of the clustering results on $154$ candidate locations that are located in the area with latitude in $47^\circ N \sim 49^\circ N$ and longitude in $1^\circ W \sim 6^\circ E$. 
By using the hierarchical clustering method, we divide these locations into $5$ clusters. We note several properties of our location clusters. First, instead of dividing areas with equal-sized grid cells~\cite{Grid_1}\cite{Grid_2}, the hierarchical clustering method only considers the user-generated locations while the areas that no user mentions are out of consideration. Second, the densities inside the clusters are different; however, the average distances between all the candidate locations in any two neighboring clusters are equal ($100km$ in Figure~\ref{fig:cluster}). Third, the complexity of the algorithm is $O(|\mathcal{L}|^3)$, where $|\mathcal{L}|$ is the total number of the candidate locations.

\begin{algorithm}[t]
\SetAlgoNoLine
\KwIn{All the candidate locations $l \in \mathcal{L}$\;}
\KwOut{Location clusters set $\mathcal{C} = \{c_1, c_2, \cdots, c_s\}$ ($s$ is the number of clusters)\;}
\textbf{\textit{Step 1}}: treat all $l \in \mathcal{L}$ as a cluster and calculate the distance between any two locations\;
\Repeat{all the candidate locations are organized into one cluster tree}{
\textbf{\textit{Step 2}}: find and merge the two closest location clusters into a new location cluster\;
\textbf{\textit{Step 3}}: compute the average distance between the new cluster and each of the old ones\;
}
\textbf{\textit{Step 4}}: cut the cluster tree into clusters with an ideal distance threshold
\caption{Clustering Locations}
\label{alg:cluster}
\end{algorithm}

\subsection{Cluster Selector}
Given a location cluster and a LN-user's location probability vector obtained by \textit{PFLI} model, we sum up the user's probabilities of locations inside the cluster as the cluster probability. Cluster selector calculates the probabilities of all the clusters that the LN-user may reside in and then selects the cluster with the highest probability.



\subsection{Location Selector}
\label{sec:L_slt} Finally, we select a best point from
the selected cluster as the user's predicted location of the current city. Three alternatives are considered. First, we select the \textit{point of the highest probability} inside the selected cluster
as the best point. Second, we consider the \textit{geographic
centroid} of the selected cluster as the user's best point. The
geographic centroid is the average coordinate for all the points in
a cluster while the probability of each point is considered as its
weight. Third, we calculate the \textit{center of minimum distance}
which has the minimum overall distance from itself to all the rest of
locations in a cluster. We will further discuss and compare the three methods in Sec.~\ref{sec:Eva}.

\subsection{Implementation of Prediction Approach}

We summarize the current city prediction approach in Algorithm~\ref{alg:ccp}. In practice, to speed up the computation of location probability vector for a given LN-user $u$, we first compute location indications from $u$'s location sensitive attributes and LN-friends:
\begin{equation}
\begin{split}
\label{pro_init}
\mathbf{p}(u)= \sum_{a_k \in \mathcal{A}} (\mu_k R_{\cdot a_k(u)} \ + \lambda_{\alpha} \sum_{v \in \mathcal{F}^{^{LN}}} \alpha_k R_{\cdot a_k(v)})
\end{split}
\end{equation}

Assume $\mathcal{L}_{_{LA-F}}$ is the set of current cities of $u$'s LA-friends. We sum location indications from $u$'s LA-friends $p_{_{LA-F}}(u,l_i)$ (refer to Eq.\ref{eq:LA-FLI}) to $p(u,l_i)$, where $l_i \in \mathcal{L}_{_{LA-F}}$.

\begin{algorithm}[t]
\SetAlgoNoLine
\KwIn{A LN-user $u$'s location sensitive attributes\; $u$'s friends list and friends' location sensitive attributes\; Location clusters set $\mathcal{C} = \{c_1, c_2, \cdots, c_s\}$ ($s$ is the number of clusters)\;}
\KwOut{Predicted current city for $u$: $\langle \textit{lat},\textit{lon}\rangle$\;}
Compute location indications $\mathbf{p}(u)$ by $u$'s location sensitive attributes and LN-friends \textbf{\textit{(Eq. \ref{pro_init})}}\;
Obtain all of LA-friends' current city $\mathcal{L}_{_{LA-F}}$\;
\For{$l_i \in \mathcal{L}_{_{LA-F}}$}{
$p(u,l_i)\leftarrow p(u,l_i)+p_{_{LA-F}}(u,l_i)$\;
}
\For{$c_x \in \mathcal{C}$}{
$p(u)_{c_x}= \sum_{l\in c_x}p(u,l)$
}
Cluster selection: $c_h$ where $p(u)_{c_h} \geq p(u)_{c_x}, \forall c_x \in \mathcal{C}$\;
Location selection from $c_h$ \textbf{\textit{(Sec. \ref{sec:L_slt})}}\;
The predicted current city of $u$: $\langle \textit{lat},\textit{lon}\rangle $
\caption{Current City Prediction}
\label{alg:ccp}
\end{algorithm}

\section{Evaluation on Current City Prediction}
\label{sec:Eva} In this section, we first introduce the experiment setups including the used Facebook data
set, the compared approaches and the measurements. Then, we report
the experiment results.

\subsection{Experiment Setup}
\subsubsection{Data description}

We crawled Facebook by a Breadth First Search (BFS)~\cite{FB_UCIrvine} approach from March to June in $2012$ and collected $371,913$ users' information including profile (e.g., gender, current city, hometown) and friends. Among all these users, $153,909$ users publicly report their current city (LA-users) and $225,314$ users do not reveal their current city (LN-users). All these users generate $12,863$ different locations. For more details about this data set, please refer to our previous work~\cite{han2015alike}.


To evaluate the prediction approach, 
a user's latest work or education experience is extracted as a location sensitive attribute, named `Work and Education'; we also exploit a user's `Hometown' as another location sensitive attribute. In our data set, $122,899$ LA-users show `Hometown', $54,097$ LA-users reveal `Work and Education' and $115,807$ LA-users publish their friend lists.

In addition to the exploited location sensitive information, some other information (e.g., a user's geo-tagged posts) in  Facebook may also leak the location. Our prediction approach can be extended to consider other location sensitive information smoothly, which we will discuss more in Sec.~\ref{subsec:extensibility_of_prediction_approach}.

\subsubsection{Approaches}
We first compare the different location selection approaches introduced in Sec.~\ref{sec:L_slt} to finalize the prediction approach with a good location selector. We also evaluate the performance of non-cluster prediction approach to show the effectiveness of location cluster. Specifically, these approaches can be denoted as:

\begin{itemize}
  \item $\textit{PFLI}_{prob}$ is a cluster based approach which selects the \textit{point of highest probability} from the selected cluster as the predicted location.
  \item $\textit{PFLI}_{cent}$ is a cluster based approach which selects the \textit{geographic centroid}\footnote{Geographic centroid is the average coordinate for all the points in a cluster while the probability of each point is considered as its weight.} from the selected cluster  as the predicted location.
  \item $\textit{PFLI}_{dist}$ is a cluster based approach which selects the \textit{center of minimum distance} from the selected cluster as the predicted location.
  \item $\textit{PFLI}_{noclst}$ is a non-cluster approach which selects the \textit{point of highest probability} from all candidate locations as the predicted location.
\end{itemize}

The proposed approaches are also compared to several state-of-the-art methods:
\begin{itemize}
  \item $\textit{Base}_{dist}$ predicts a user's location based on the observation that the distance between two users decreases by the increase of their friendship~\cite{find_me}.

  \item $\textit{Base}_{ann}$ maps any location sensitive attribute value to a certain location and applies artificial neural network to train a current city prediction model \cite{CCP}.

\item $\textit{Base}_{freq}$, borrowing the idea from the prior works based on the Twitter data set~\cite{T_content_prediction}\cite{geo_twitter}, counts the frequency of locations that emerge in a user's friends and predicts his current city by the most frequent location.

 \item $\textit{Base}_{freq+}$ improves $\textit{Base}_{freq}$ by further using the neighborhood smoothing approach~\cite{geo_twitter}. Given a location $l$, the points that are less than $20km$ apart from $l$ are considered as $l$'s neighborhoods.

  \item $\textit{Base}_{knn}$ also relies on the frequency idea for Twitter; however, it merely counts on a user's $k$ closest friends who have the most common friends with him to compute the most frequent location \cite{TweetHood}\cite{Tweecalization}.
\end{itemize}

Among the above approaches, $\textit{Base}_{dist}$ and
$\textit{Base}_{ann}$ are originally devised for Facebook; while
$\textit{Base}_{freq}$, $\textit{Base}_{freq+}$ and $\textit{Base}_{knn}$ are on Twitter. We
utilize the main ideas from $\textit{Base}_{freq}$, $\textit{Base}_{freq+}$ and
$\textit{Base}_{knn}$, and adopt them to fit our data set. By
comparing our approach to $\textit{Base}_{dist}$,
$\textit{Base}_{freq}$, $\textit{Base}_{freq+}$ and $\textit{Base}_{knn}$ which mainly depend
on friendships, we test the effectiveness of integrating location sensitive attributes. By comparing to $\textit{Base}_{ann}$, we examine the newly introduced
one-attribute/multiple-locations mapping method.

\subsubsection{Measurement}
Two widely used measurements:
\textit{Average Error Distance} (\textit{AED}) and
\textit{Accuracy within K km} (\textit{ACC@K})~\cite{T_content_prediction}\cite{geo_twitter}\cite{UID} are exploited.

\textit{Error Distance} computes the distance in kilometers between a
user $u$'s real location and predicted location, i.e., $ErrDist(u)$. \textit{AED}
averages the \textit{Error Distances} of the overall evaluated
users, denoted as $\textit{AED} = \frac{\sum_{u \in
U}ErrDist(u)}{|U|}$. In addition, we rank the users by their \textit{Error Distance} in descending order and report
\textit{AED} of the top $60\%$, $80\%$ and $100\%$ of the evaluated
users in the ranked list, denoted as $\textit{AED@}60\%$, $\textit{AED@}80\%$ and
$\textit{AED@}100\%$ respectively \cite{UID}.

Given a predefined \textit{Error Distance $K$ km}, a prediction for a user is considered as a correct prediction, if the predicted \textit{Error Distance} is less than $K$ km; otherwise, the prediction is incorrect. Then, \textit{Accuracy within K km} is defined as the percentage of correct predictions (i.e., the percentage of users being predicted with an \textit{Error Distance} less than $K$ km), denoted as $\textit{ACC@K} = \frac{|\{u|u \in U
\wedge ErrDist(u)< K\}|}{|U|}$. \textit{ACC@K} shows the
prediction capability of an approach at a specific pre-established
\textit{Error Distance}.

\subsection{Experiment Results}

\begin{table}
\centering
\scriptsize
\begin{tabular}{c||c|c|c|c|c|c|c|c|c}
  \hline
  \bfseries  & $\textit{Base}_{dist}$ & $\textit{Base}_{ann}$ & $\textit{Base}_{freq}$ & $\textit{Base}_{freq+} $ & $\textit{Base}_{knn}$ & $\textit{PFLI}_{noclst}$ & $\textit{PFLI}_{dist}$ & $\textit{PFLI}_{cent}$ & $\textit{PFLI}_{prob}$\\
  \hline
  $\textit{AED@}60\%$ & 8.6 & 5.7 & 5.9 & 4.9 & 10.8 & 2.5 & 49.5 & 5.6 & \bfseries 2.1\\
  \hline
  $\textit{AED@}80\%$ & 85.0 & 64.3 & 91.8 & 56.0 & 100.0 & 40.1 & 77.4 & 38.0 & \bfseries 36.9\\
  \hline
 $\textit{AED@}100\%$ & 1288.5 & 1129.0 & 1160.5 & 1123.7 & 1397.6 & 874.0 & 885.9 & 855.3 & \bfseries 854.4\\
  \hline
\end{tabular}
\caption{Prediction Results (\textit{AED}) for Users with LA-Friends}
\label{table:AED_UF}
\end{table}

\begin{table}
\centering
\scriptsize
\begin{tabular}{c||c|c|c|c|c|c|c|c|c}
  \hline
  \bfseries  & $\textit{Base}_{dist}$ & $\textit{Base}_{ann}$ & $\textit{Base}_{freq}$  & $\textit{Base}_{freq+} $ & $\textit{Base}_{knn}$ & $\textit{PFLI}_{noclst}$ & $\textit{PFLI}_{dist}$ & $\textit{PFLI}_{cent}$ & $\textit{PFLI}_{prob}$\\
  \hline
  $\textit{AED@}60\%$ & 102.8 & 6.7 & 73.9 & 66.6 & 119.5 & 3.5 & 50.6 & 6.3 & \bfseries 3.1\\
  \hline
  $\textit{AED@}80\%$ & 1368.8 & 74.7 & 1257.2 & 1243.1 & 1429.6 & 52.5 & 88.2 & 50.2 & \bfseries 49.1\\
  \hline
 $\textit{AED@}100\%$ & 2671 & 1204.0 & 2523.5 & 2498 & 2698.5 & 981.0 & 989.9 & 960.8& \bfseries 960.0\\
  \hline
\end{tabular}
\caption{Prediction Results (\textit{AED}) for Overall Users}
\label{table:AED_Uall}
\end{table}

Many relationship-based methods (e.g., $\textit{Base}_{dist}$, $\textit{Base}_{freq}$, $\textit{Base}_{freq+}$ and $\textit{Base}_{knn}$) rely heavily on users' LA-friends whose locations are exposed. In general, such methods can work well for the users who have a certain number of LA-friends; but when they are applied to the overall users (who either have or do not have LA-friends), the performance notably decreases. We evaluate the prediction performance on two user sets: \textit{users with LA-friends} and \textit{overall users}, and report the evaluation results on \textit{AED} and \textit{ACC@K} subsequently.

\subsubsection{Evaluation on \textit{AED}}
Table \ref{table:AED_UF} and Table \ref{table:AED_Uall} show the \textit{AED}s of all the compared approaches for two user sets. The smallest \textit{AEDs}, which are generated by $\textit{PFLI}_{prob}$, have been highlighted in bold.

Let us first look at the \textit{PFLI} model based approaches (i.e., $\textit{PFLI}_{dist}$, $\textit{PFLI}_{cent}$, $\textit{PFLI}_{prob}$, and $\textit{PFLI}_{noclst}$). Among the first three cluster based approaches that are different at their location selectors, $\textit{PFLI}_{dist}$ generates the largest \textit{AEDs} while $\textit{PFLI}_{prob}$ achieves the smallest \textit{AEDs}. We also compare the non-cluster approach $\textit{PFLI}_{noclst}$ and the cluster approach $\textit{PFLI}_{prob}$, which both select location of the highest probability. We observe that $\textit{PFLI}_{prob}$ presents smaller \textit{AEDs} than $\textit{PFLI}_{noclst}$ and verify the effectiveness of the location cluster approach.

In addition, the results show that the \textit{PFLI} model based approaches present much smaller \textit{AEDs} than all the other baselines. In particular, the results demonstrate the \textit{PFLI} model based approaches mapping one-attribute to multiple locations reduce the \textit{AED} significantly compared to $\textit{Base}_{ann}$ which maps one-attribute to one-location.

By examining the results of $\textit{AED@60\%}$, $\textit{AED@80\%}$ and $\textit{AED@100\%}$, we observe that the \textit{PFLI} model based approaches can predict current city with relatively small $\textit{AED@60\%}$ and $\textit{AED@80\%}$; whereas, $\textit{AED@100\%}$ increases by $10$--$23$ times from $\textit{AED@80\%}$. This demonstrates the large \textit{Error Distance} only occurs at predictions for a small number of users.

Lastly, we compare the results in the two Tables and notice that the prior approaches ($\textit{Base}_{dist}$, $\textit{Base}_{freq}$, $\textit{Base}_{freq+}$ and $\textit{Base}_{knn}$) predict locations with much larger \textit{AEDs} for \textit{overall users} than for \textit{users with LA-friends}; however, for the \textit{PFLI} model based approaches,  \textit{AEDs} differ slightly for two user sets. It demonstrates that a user's profile can significantly contribute to the location prediction when the user's friends' locations are unavailable.

\subsubsection{Evaluation on \textit{ACC@K}}

\begin{figure}[!htbp]
\centering \subfigure[Users with LA-Friends]{
\includegraphics[width=5.5cm, height=5cm]{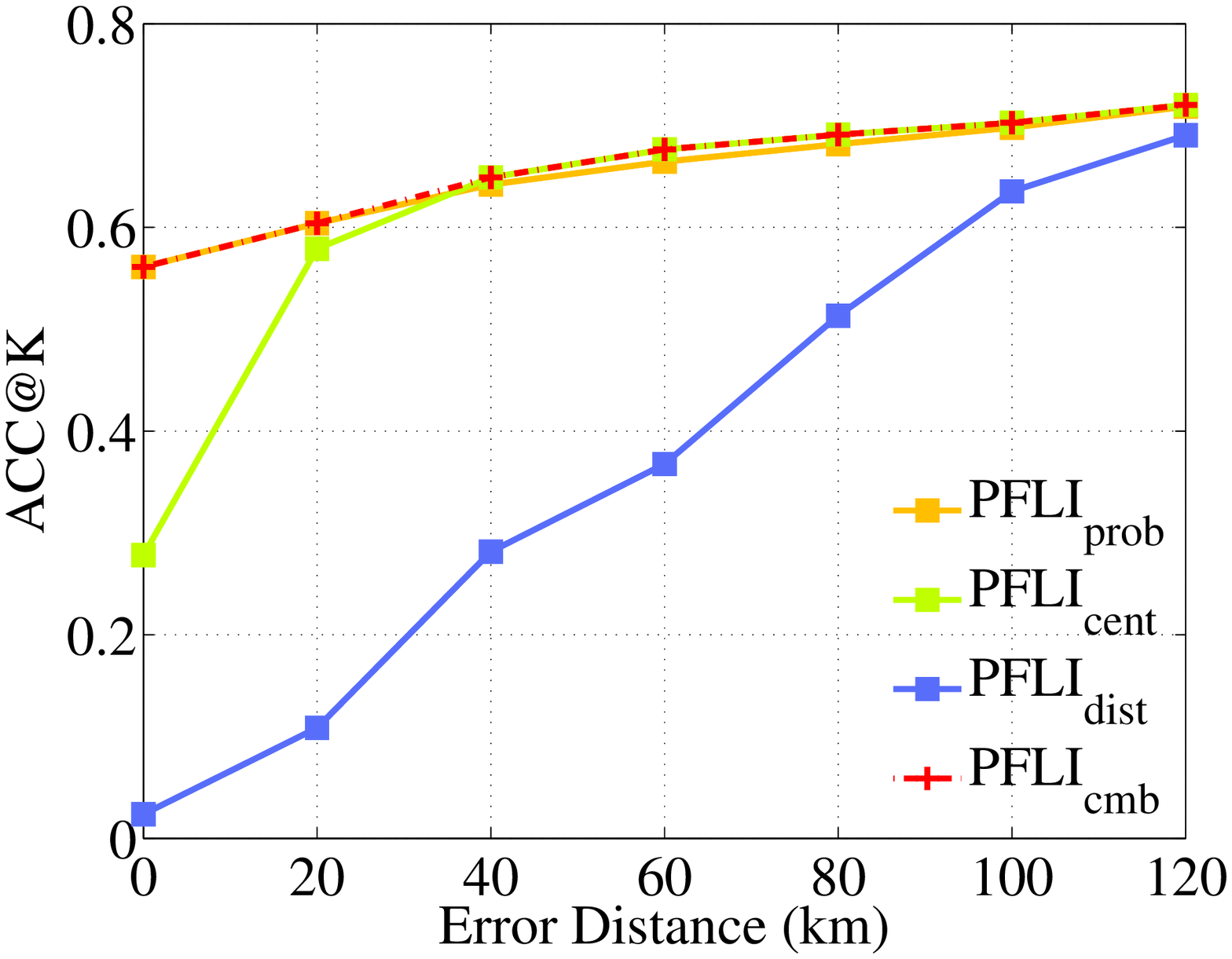}
\label{fig:lb_slt} } \subfigure[Overall Users]{
\includegraphics[width=5.5cm, height=5cm]{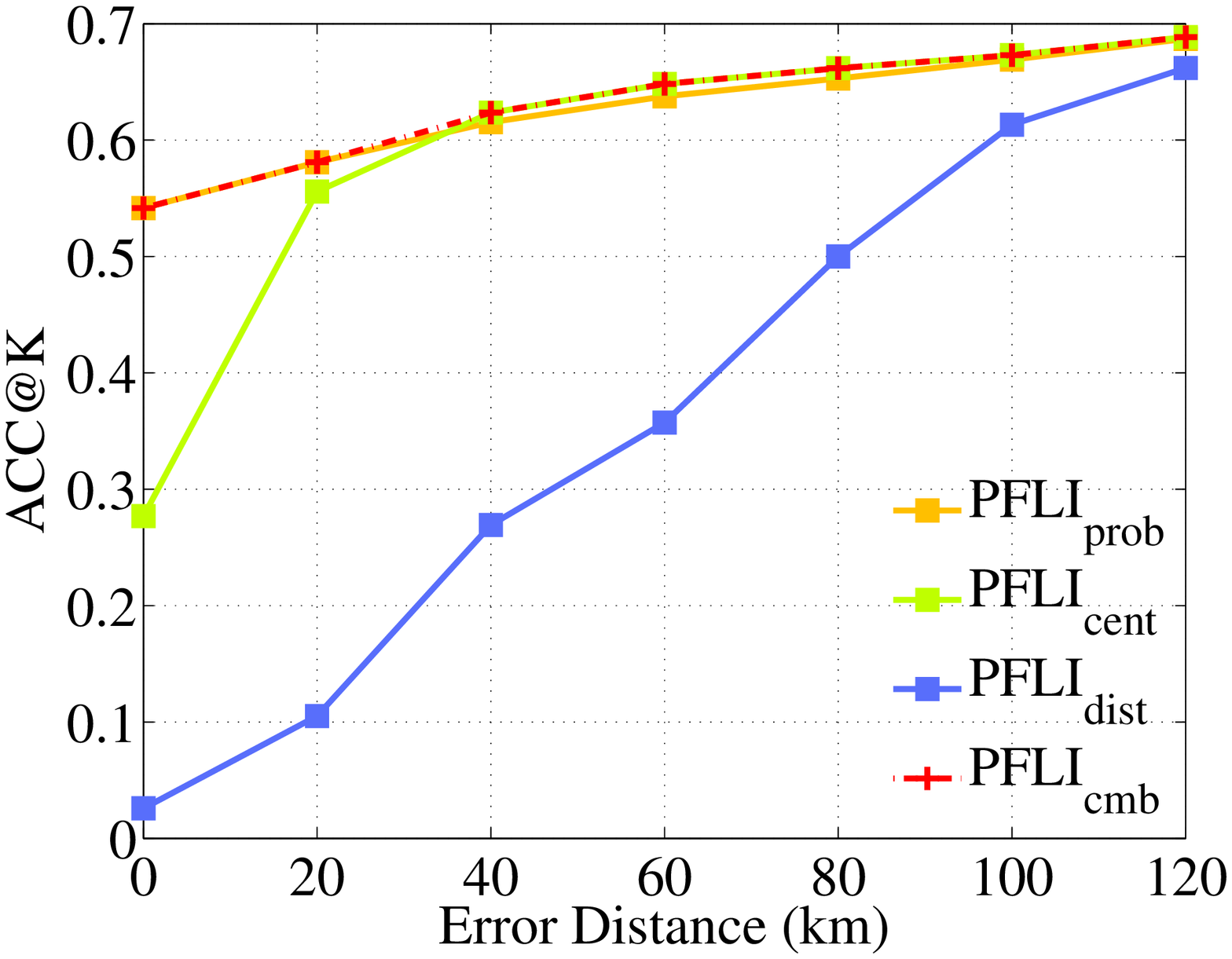}
\label{fig:slt} }\caption{\textit{ACC@K} of Different
Location Selectors.} \label{fig:Selt}
\end{figure}

We study \textit{ACC@K} of the three proposed prediction approaches ($\textit{PFLI}_{prob}$, $\textit{PFLI}_{cent}$ and $\textit{PFLI}_{dist}$) for two user sets in Figure \ref{fig:Selt}. We observe that the accuracy of $\textit{PFLI}_{prob}$ goes up steadily with the increase of \textit{Error Distance}. $\textit{PFLI}_{cent}$ may lead to very low accuracy when the pre-established \textit{Error Distance} is quite small; but it can achieve higher accuracy than $\textit{PFLI}_{prob}$, when the pre-established \textit{Error Distance} is larger than $40$ km. This reveals the properties of these two prediction approaches: $\textit{PFLI}_{cent}$, which selects the geographic centroid of a cluster, generates a short average \textit{Error Distance} to all the locations in the cluster but fails to pick the user's exact coordinate once it is not the centroid; while $\textit{PFLI}_{prob}$ may produce a large \textit{Error Distance} if the location of the highest probability is not the user's real location. In addition, $\textit{PFLI}_{dist}$ is not competitive with the other two approaches.

Rather than solely using any one of the proposed approaches, we exploit a combined-approach strategy by flexibly selecting the best approach according to the pre-established \textit{Error Distance}. Specifically, this strategy uses $\textit{PFLI}_{prob}$ when the pre-established \textit{Error Distance} is smaller than $40$ km and otherwise applies $\textit{PFLI}_{cent}$. The combination is practical and can obtain a better performance than using any single approach. We plot the combination line in Figure \ref{fig:Selt} and call it $\textit{PFLI}_{cmb}$.

\begin{figure}[!htbp]
\centering \subfigure[Users with LA-Friends]{
\includegraphics[width=5.5cm, height=5cm]{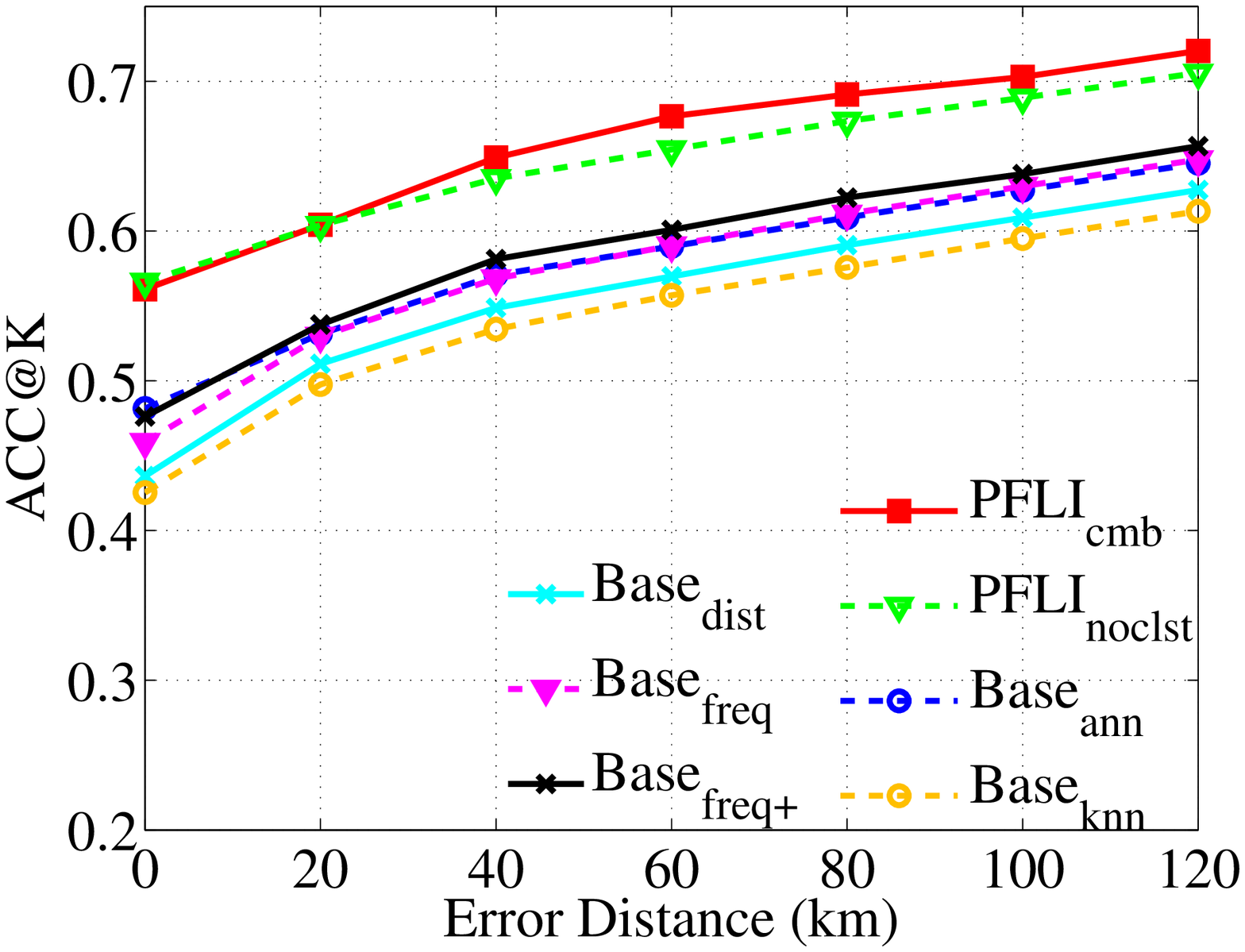}
\label{fig:lb_acc} } \subfigure[Overall Users]{
\includegraphics[width=5.5cm, height=5cm]{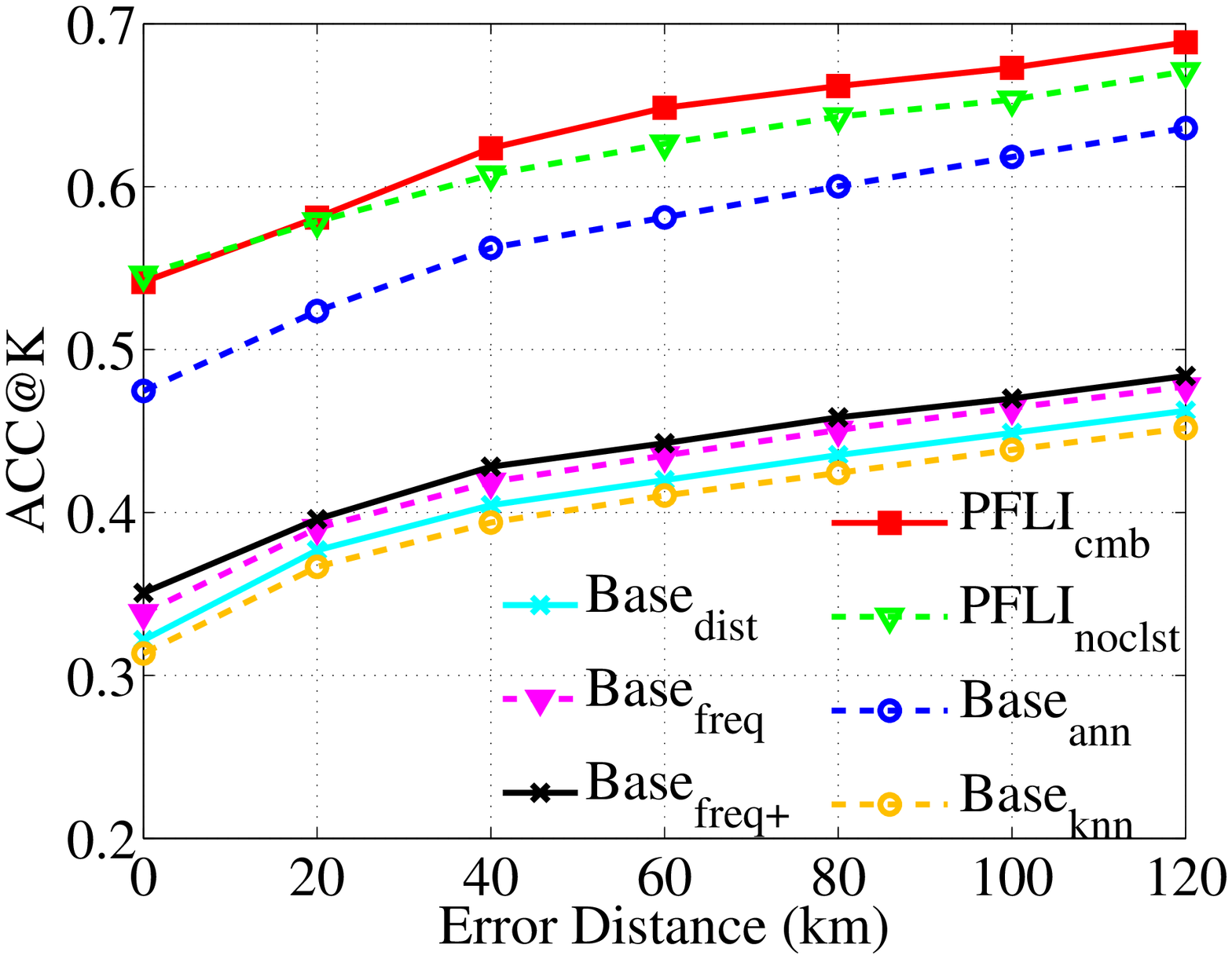}
\label{fig:acc} } \caption{\textit{ACC@K} of the Proposed
Approach and Other Baselines.}
\label{fig:ACC}
\end{figure}

Figure \ref{fig:ACC} compares $\textit{PFLI}_{cmb}$ to various baseline methods in terms of \textit{ACC@K}. We observe that the proposed $\textit{PFLI}_{cmb}$ outperforms all the compared baselines for both user sets. Compared to $\textit{PFLI}_{noclst}$, $\textit{PFLI}_{cmb}$ increases around $1.5\%$ and $1.2\%$ of accuracy on average for \textit{users with LA-friends} and \textit{overall users}. This proves the effectiveness of the cluster strategy with successive cluster selection and location selection.

Comparing Figure~\ref{fig:lb_acc} and \ref{fig:acc}, we observe that the approaches $\textit{Base}_{freq}$, $\textit{Base}_{freq+}$, $\textit{Base}_{dist}$ and $\textit{Base}_{knn}$ perform much worse for \textit{overall users} than for \textit{users with LA-friends}. This observation again indicates that these approaches depend heavily on the friends' locations. However, in respect of the other approaches, which integrate location indications from both location sensitive attributes and friends (including our previous work $\textit{Base}_{ann}$~\cite{CCP}), the prediction performance for \textit{overall users} relatively approaches to the performance for \textit{users with LA-friends}.

\section{Current City Exposure Estimator}
\label{sec:exposure} In this section, we pay attention to estimating current city exposure probability for a user who hides his
current city. We formulate the \textbf\textit{{current city
exposure estimation}} problem as: \textit{Given, (i)~a graph $\mathcal{G}=(\mathcal{U}^{^{LA}}\cup \mathcal{U}^{^{LN}},\mathcal{E},\mathcal{L})$; (ii)~the public location $l(u)$ for LA-users $u \in \mathcal{U}^{^{LA}}$; (iii)~the location sensitive attributes $\mathcal{A}(u)$ and the friends list $\mathcal{F}(u)$ for all the users $u \in (\mathcal{U}^{^{LA}} \cup \mathcal{U}^{^{LN}})$; $(iv)$ a  pre-established Error Distance $K$ km, we forecast the current city exposure probability within $K$
km and report the exposure risk level for each LN-user $u \in \mathcal{U}^{^{LN}}$}.

To solve this problem, we run the proposed prediction approach on an
aggregation of users and conduct analysis on the aggregated
prediction results. Furthermore, we apply a regression method to
construct the exposure model according to the analysis observations.
Relying on this model, we devise a current city exposure estimator
to inform users of their current city \textit{Exposure Probability within
$K$ km} and \textit{Exposure Risk Level}.

The \textit{Exposure Probability within $K$ km (EP@K)} represents
the probability that a user's current city could be inferred
correctly if the pre-established \textit{Error Distance} is $K$ km. As it
is conceptually similar to the metric \textit{ACC@K}, we compute
it by the same formula:
\begin{equation}
\label{eq:exp_pro}
EP@K = \frac{|\{u|u \in U \wedge ErrDist(u)<K\}|}{|U|}
\end{equation}

Additionally, we set up five \textit{Exposure Risk Levels} according
to the value of \textit{Exposure Probability}, shown in Table
\ref{table:Risk_Level}. \textit{Level 5} is defined as the most risky
level, which indicates an \textit{Exposure Probability} higher than
$0.9$, while \textit{Level 1} is the safest one, which represents a
small \textit{Exposure Probability} lower than $0.25$.

Next, we show some observations of inspections on the aggregated
prediction results. We then introduce the current
city exposure model and the model based estimator. Finally, we
illustrate some case studies to show the use of our proposed
exposure estimator. We also summarize some guidelines to reduce the
exposure risk.

\begin{table}
\centering
\scriptsize
\begin{tabular}{c||c|c|c|c|c}
  \hline
  \bfseries Exposure Probability & $[0.9,1]$ & $[0.75,0.9)$ & $[0.5,0.75)$ & $[0.5,0.25)$ & $ [0.25,0]$ \\
  \hline
  \bfseries Risk Level & Level 5 & Level 4 & Level 3 & Level 2 & Level 1\\
  \hline
\end{tabular}
\caption{Risk Level vs. Exposure Probability}
\label{table:Risk_Level}
\end{table}

\subsection{Current City Exposure Inspection}
\label{sec:expo_inspection}
In this subsection, we extract several measurable characteristics from users' self-exposed information (e.g., User Category), and inspect the current city exposure probability by these characteristics.


First, we classify users into diverse categories with respect to the
combinations of visible/invisible properties of their location
sensitive attributes and friends list. Table \ref{table:category}
lists the obtained seven \textit{User Categories}. \textit{User
Category} measures the types and amount of users' self-exposed
information.

\begin{table}
\centering
\scriptsize
\begin{tabular}{c|c}
  \hline
 \bfseries User's Visible Attributes & \bfseries Abbreviation\\
  \hline
`Hometown' & `HT'\\
`Work and Education' & `WE'\\
`Friends' & `F'\\
`Hometown' and `Work and Education' & `HT+WE'\\
`Hometown' and `Friends' & `HT+F'\\
`Work and Education' and `Friends' & `WE+F'\\
`Hometown', `Work and Education' and `Friends' & `HT+WE+F'\\
  \hline
\end{tabular}
\caption{Users Categories by Visible Attributes Combination}
\label{table:category}
\end{table}

\begin{figure}[!htp] \centering
\includegraphics[width=6cm, height=5.5cm]{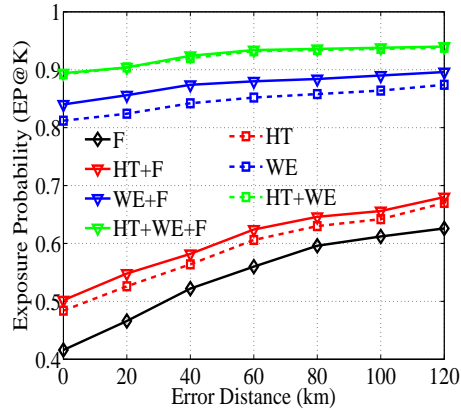}
\caption{Current City Exposure Probability by User Category.}
\label{fig:Exposure_category}
\end{figure}
Figure \ref{fig:Exposure_category} inspects the \textit{Exposure
Probabilities} for various \textit{User Categories}. From this
figure, we observe that different types of self-exposed information
may divulge users' current city to different extent. For instance,
users in `WE' category are normally more dangerous to disclose their
current city than users in `HT' or `F' categories. We also find that
the users who publish their `WE' (in `WE', `HT+WE', `WE+F'
or `HT+WE+F' categories) exhibit a high \textit{Exposure Probability}. This
means that `WE' is a very risky attribute to leak users' current
city. The results also reveal that `HT' is more sensitive to
disclose current city than `F', although `F' is generally regarded
as a significant location indication.

Figure \ref{fig:Exposure_category} also indicates that a user's current city generally could be predicted with a higher
probability if the user exposes more information. For example, users
who expose `HT+F' exhibit a higher exposure probability than users
only revealing either `HT' or `F'. Note that, for a user who exposes
`HT+WE', his current city exposure probability can be up to $90\%$,
which approaches to the exposure probability of users who expose
`HT+WE+F'. In other words, merely exposing `HT+WE' can
almost lead to the exposure of a user's current city. To conclude, \textit{User
Category}, which distinguishes users by the types and amount of their
self-exposed information, relates to \textit{Exposure Probability}.


In addition to \textit{User Category}, we study the influence of the percentage of friends with attributes (i.e., \textit{\%~Friends with Attributes}) on \textit{Exposure Probability}. \textit{\%~Friends with Attributes} is the ratio of a user's friends who present at least one attribute to his overall friends.

\begin{figure}[!htp] \centering
\includegraphics[width=6cm, height=5.5cm]{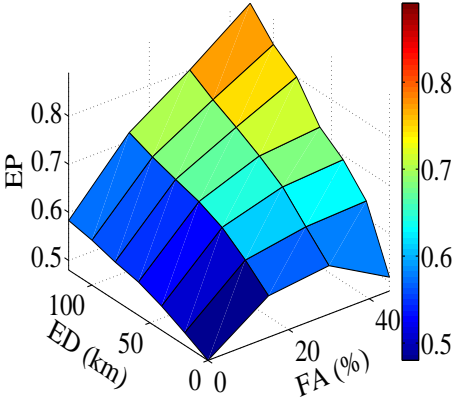}
\caption{Current City Exposure Probability by the Percentage of Friends with Attributes.}
\label{fig:Exposure_PF}
\end{figure}

Figure \ref{fig:Exposure_PF} displays the \textit{Exposure Probability} (i.e., EP, $Z$ axis) by \textit{\% Friends with Attributes} (i.e., FA, $X$ axis) at different \textit{Error Distances} (i.e., ED, $Y$ axis). As more than $95\%$ of the users have a \textit{\% Friends with Attributes} smaller than $45\%$, we only look at its value in a range of $0\%$ to $45\%$.
Generally speaking, \textit{Exposure Probability} grows by the increase of \textit{\%~Friends with Attributes}.

\begin{figure*}[!htb]
\centering \subfigure[`HT']{
\includegraphics[width=3cm, height=3cm]{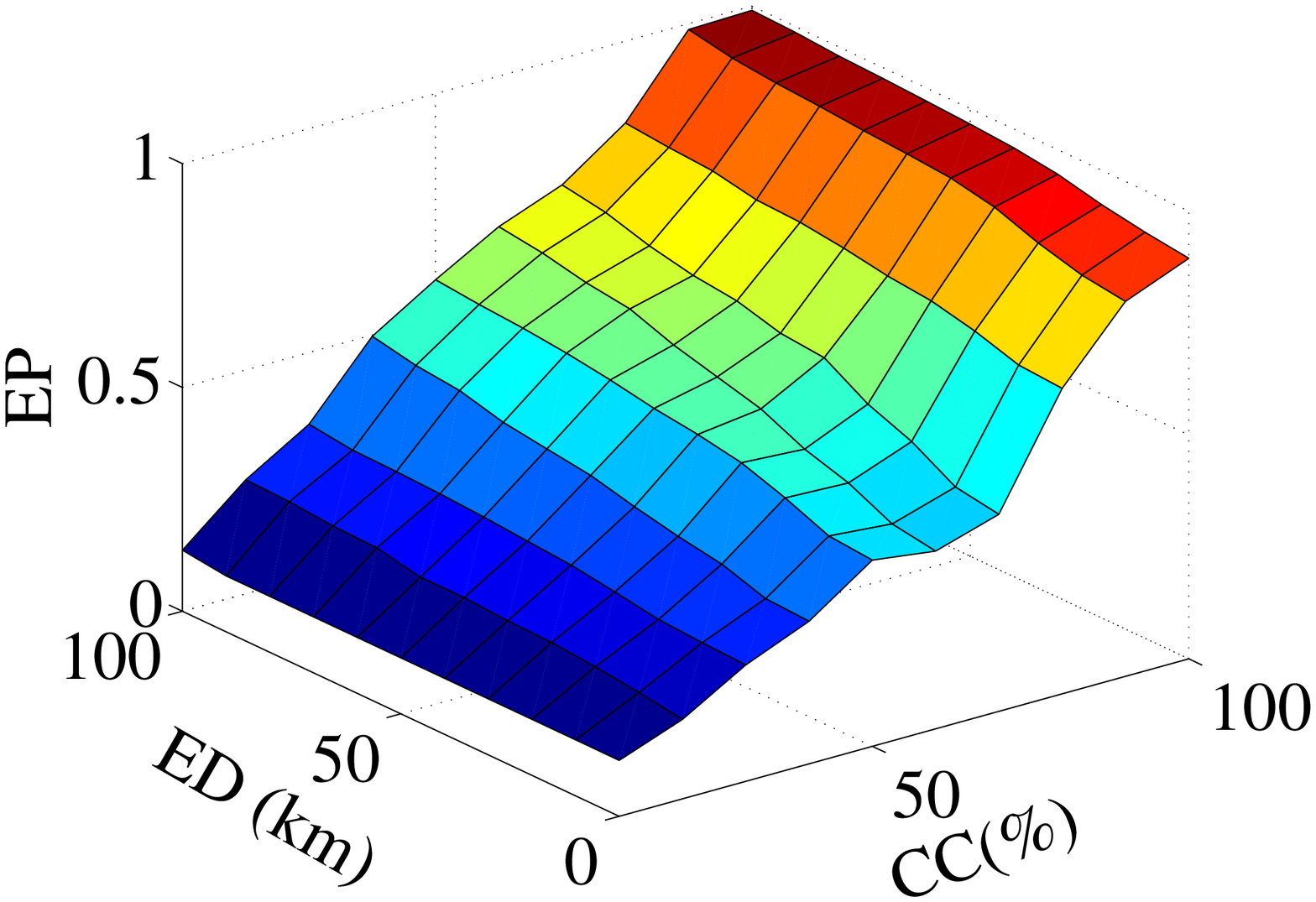}
\label{fig:ht} } \subfigure[`WE']{
\includegraphics[width=3cm, height=3cm]{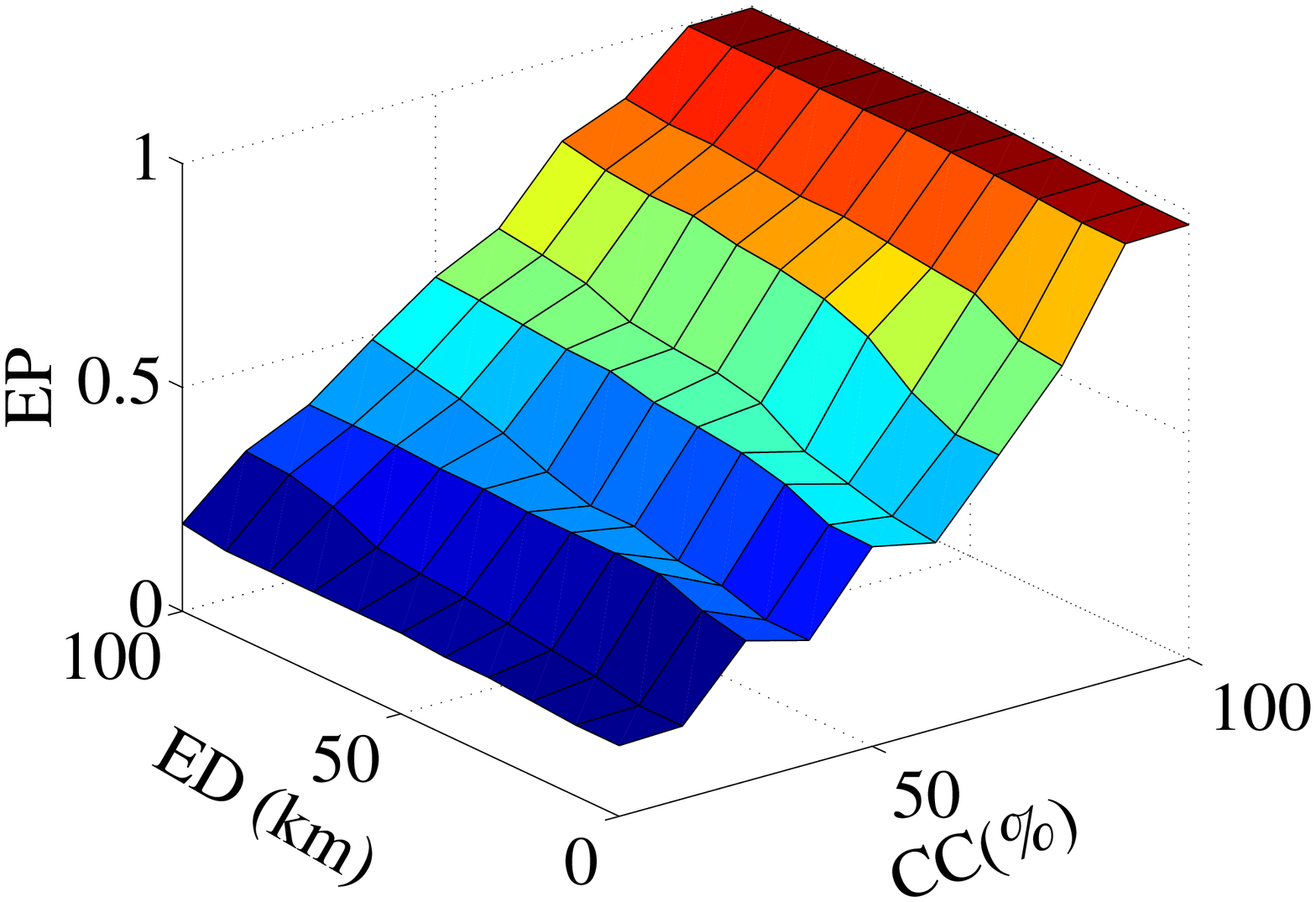}
\label{fig:em} } \subfigure[`F']{
\includegraphics[width=3cm, height=3cm]{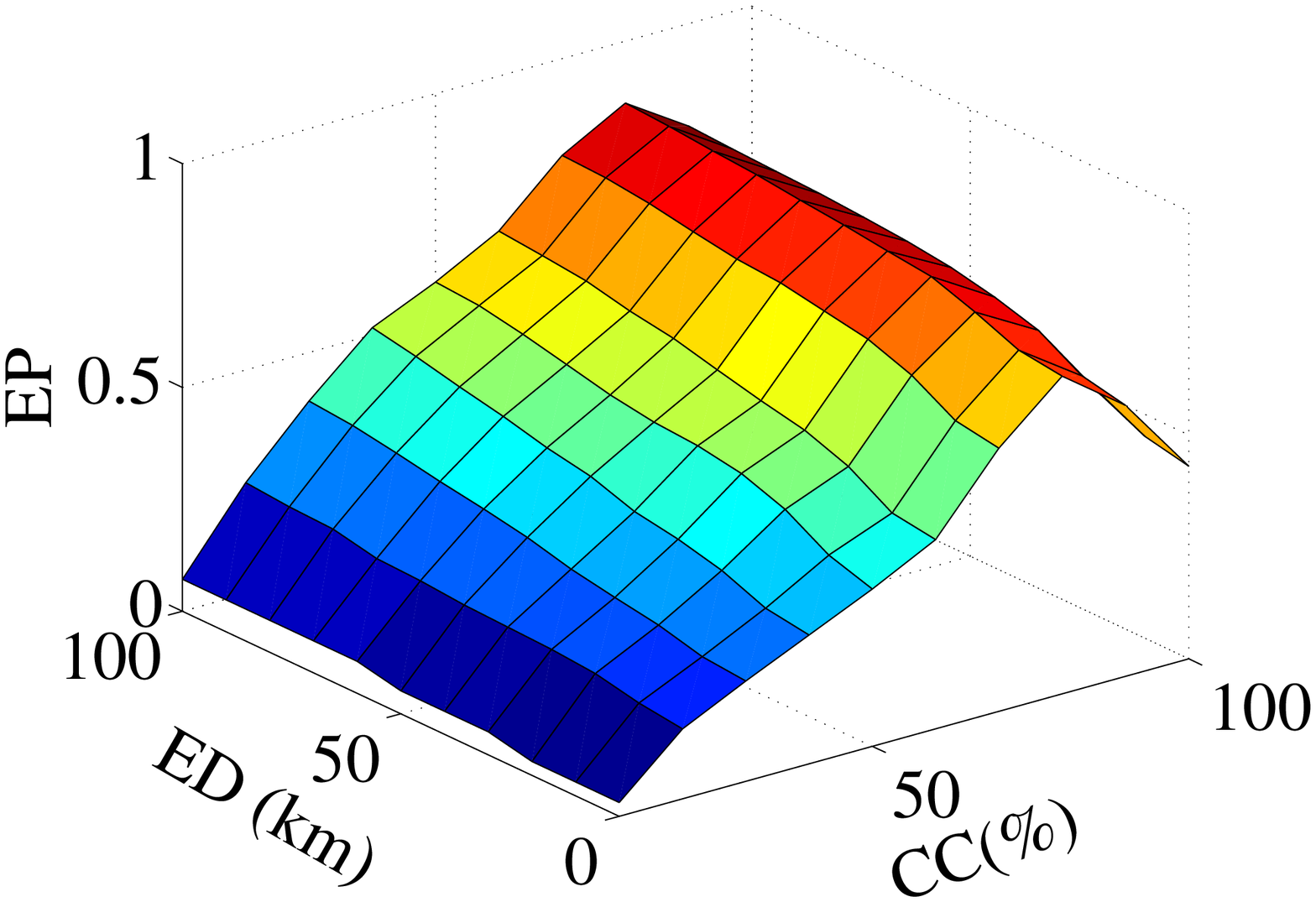}
\label{fig:f} } \subfigure[`HT+WE']{
\includegraphics[width=3cm, height=3cm]{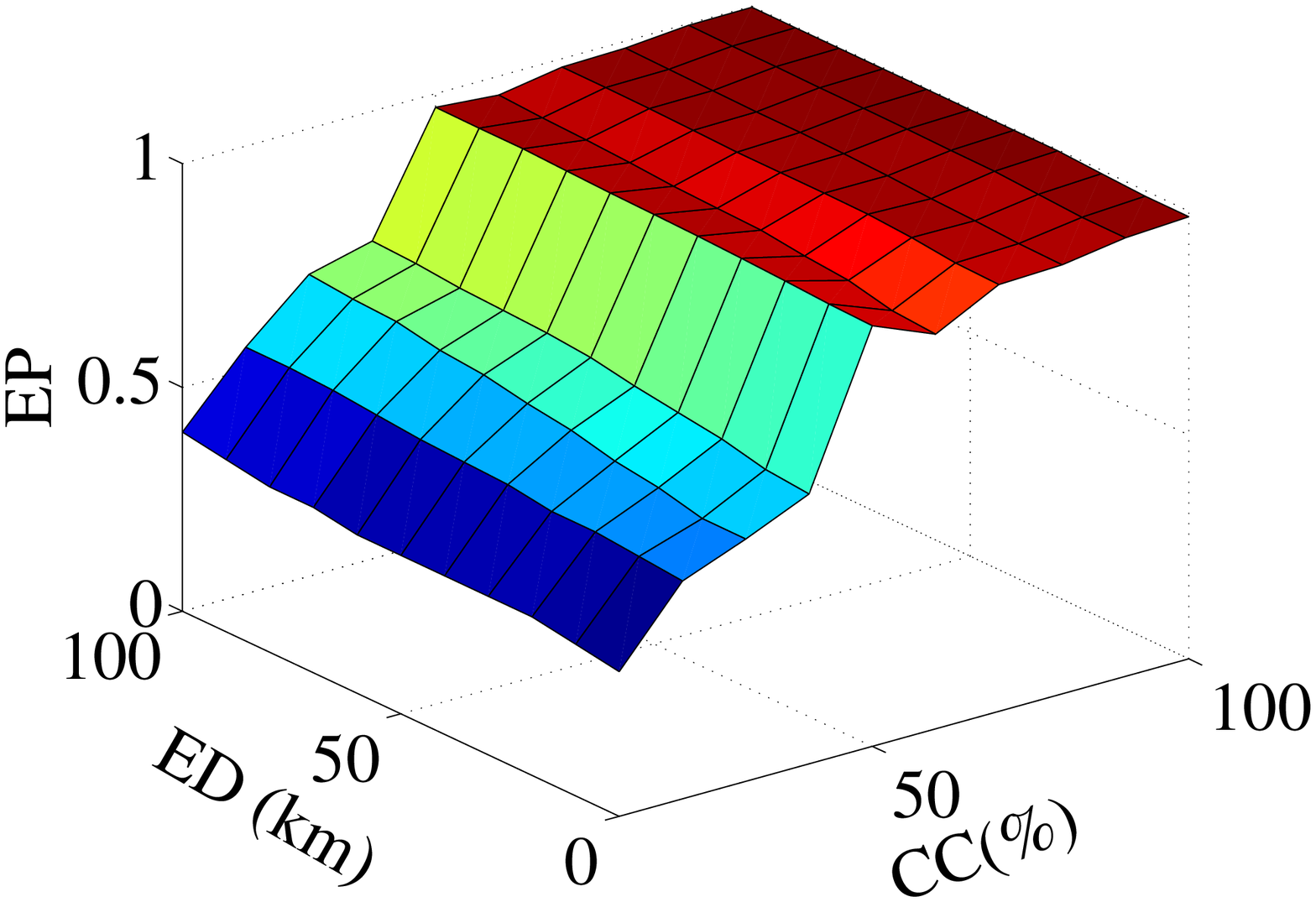}
\label{fig:htem} } \subfigure[`HT+F']{
\includegraphics[width=3cm, height=3cm]{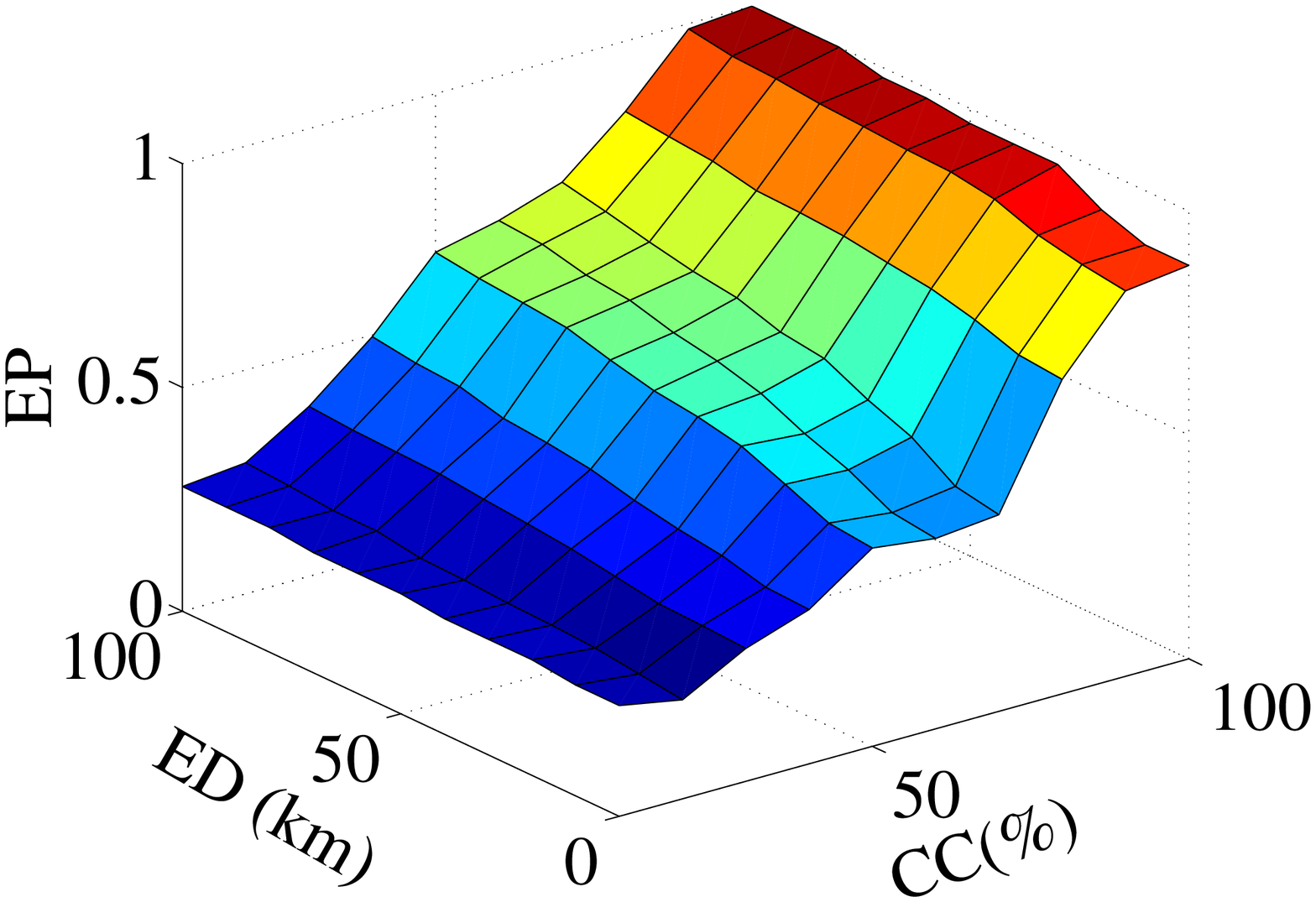}
\label{fig:htf} } \subfigure[`WE+F']{
\includegraphics[width=3cm, height=3cm]{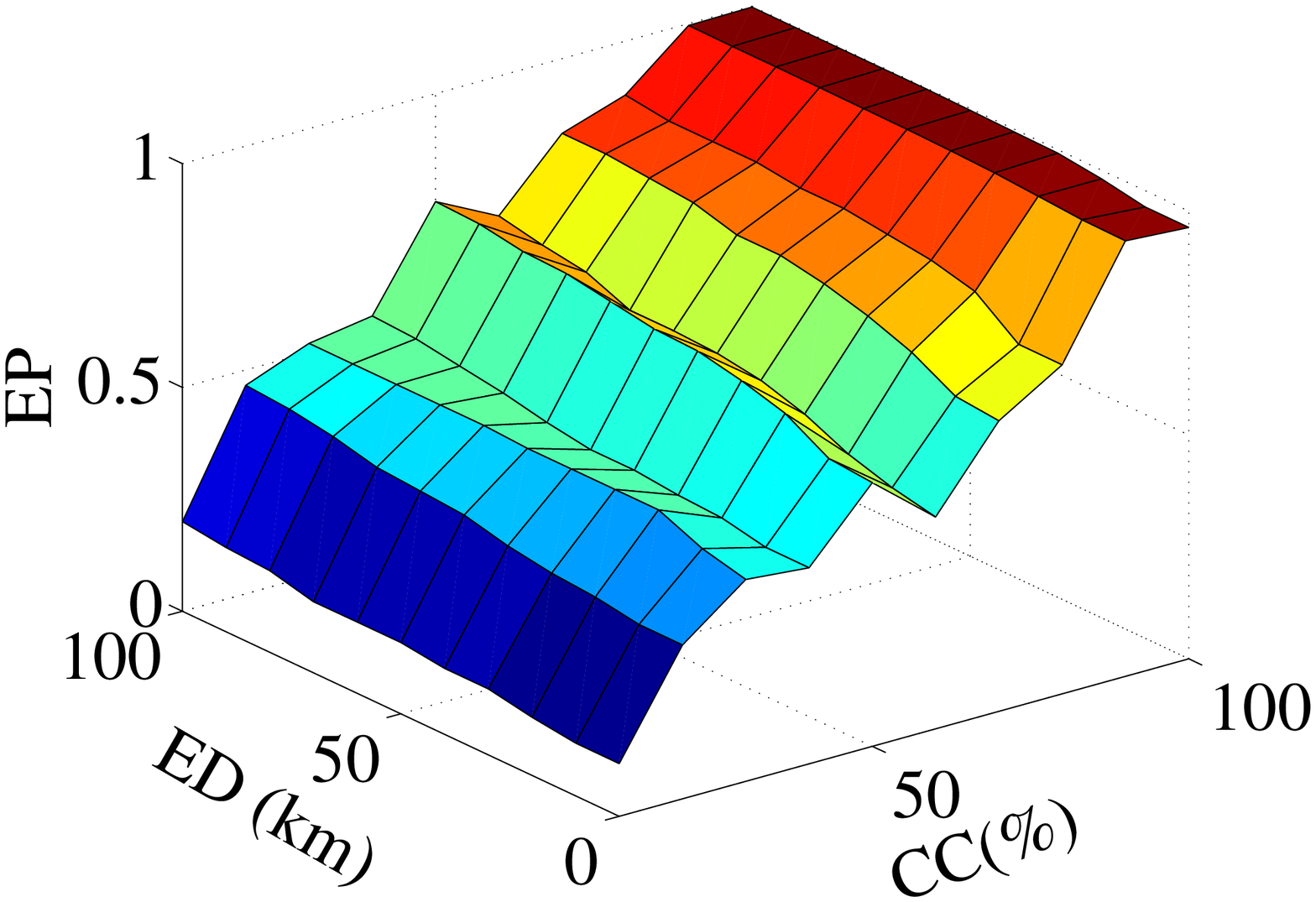}
\label{fig:emf} } \subfigure[`HT+WE+F']{
\includegraphics[width=3.3cm, height=3cm]{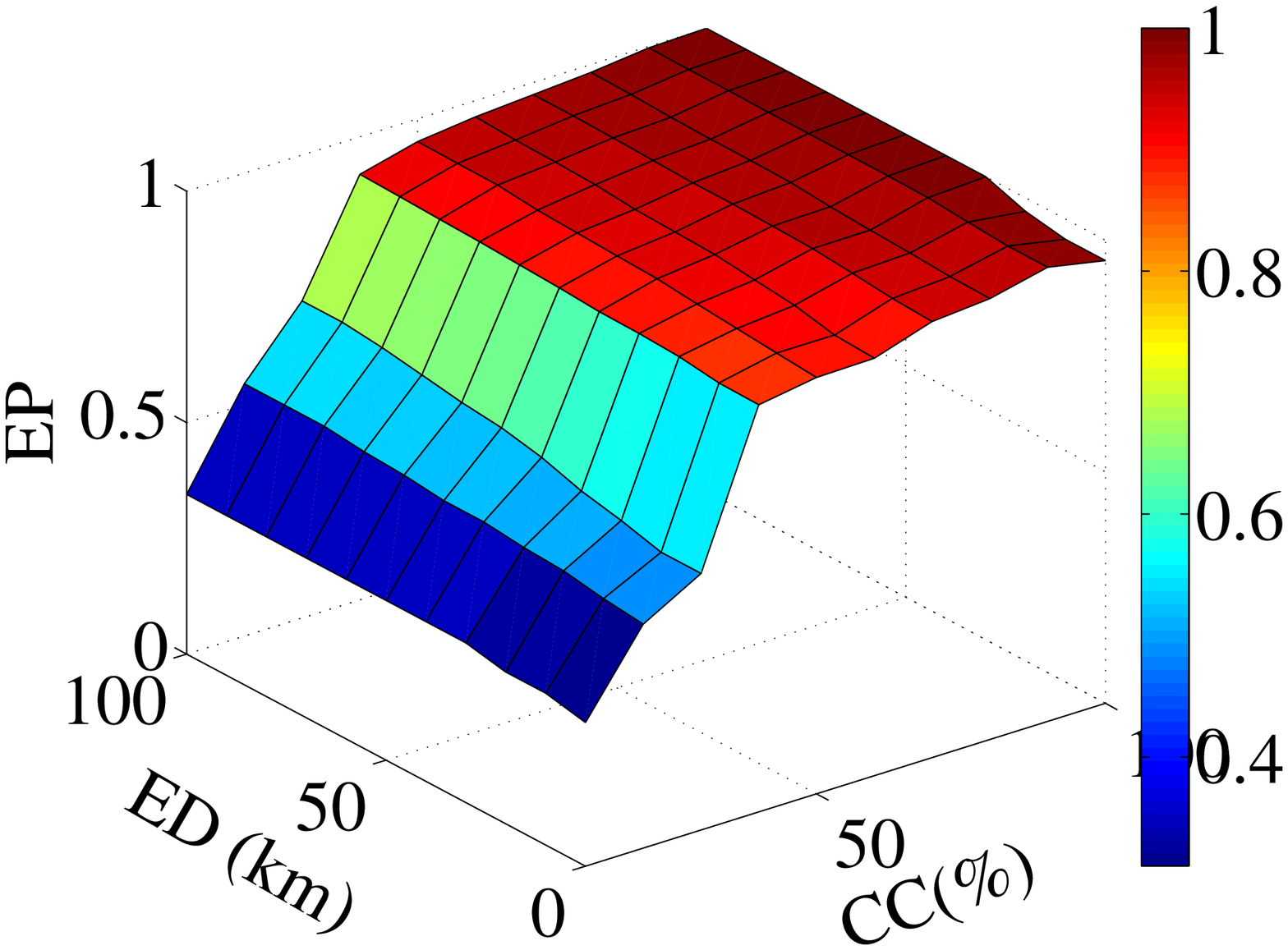}
\label{fig:htemf} } \caption{Exposure Probability by Cluster Confidence in Different User Categories.}
\label{fig:Exposure_Coefficient}
\end{figure*}

In addition, we define a new metric named \textit{Cluster Confidence}. It estimates the ratio of the
probabilities of candidate locations in the selected cluster $c_h$
to the overall probabilities of all the candidate locations (equal
$1$), calculated as follows:
\begin{equation}
\label{eq:cluster_confidence}
CC(u) = \frac{\sum_{l\in c_h}p(u,l)}{\sum_{l\in \mathcal{L}}p(u,l) } = \sum_{l\in c_h}p(u,l)
\end{equation}

\textit{Cluster Confidence} represents the confidence of the users' location indications. For example, \textit{Cluster Confidence} with a value of $100\%$ means that all of a user's
location indications point to an exclusive location cluster. We
further look into the change of \textit{Exposure Probability} according to
\textit{Cluster Confidence} for each \textit{User Category}.

Figure \ref{fig:Exposure_Coefficient} reveals how \textit{Exposure
Probability} (i.e., EP, $Z$ axis) varies with diverse \textit{Cluster Confidence} (i.e., CC, $X$ axis) and
\textit{Error Distances} (i.e., ED, $Y$ axis) in different \textit{User Categories}. The results show that the \textit{Exposure Probability} normally grows up when the
\textit{Cluster Confidence} gets larger. When the \textit{Cluster Confidence} equals $100\%$, the \textit{Exposure Probability}
surpasses $90\%$ within a pre-established \textit{Error Distance} of $20$
km almost for all \textit{User Categories}. This observation
indicates that the current city is more dangerous to be predicted
when a user's location indications are more likely to point to one
city or to multiple cities that are in the same cluster. In other
words, a user's current city can be easily disclosed if the confidence
of the user's self-exposed information is high.

Note that, there exists an exception for the users only exposing
their `F': the decline of \textit{Exposure Probability} when the
\textit{Cluster Confidence} is larger than $0.9$. One reasonable
explanation is that only the users with an extremely small number of
friends (e.g., only one friend) can have the \textit{Cluster Confidence} higher than $0.9$, which might reduce the exposure risk of
current city due to the limited information.

\subsection{Estimating Current City Exposure Risk}
\label{expo_extimation}
\subsubsection{Current City Exposure Model}
In the previous section, we observe that a user's current city
\textit{Exposure Probability} is probably influenced by four
factors: \textit{Error Distance}, \textit{User Category}, \textit{\% Friends with Attributes} and
\textit{Cluster Confidence}. Taking these four factors as features, we respectively use Random Decision Forest and Linear Regression approaches to model \textit{Exposure Probability}. The performance of model is evaluated by two commonly used metrics, \textit{Mean Absolute Error} (MAE) and \textit{Root Mean Squared Error} (RMSE), with $10$-cross validation, shown in Table~\ref{table:EM_comp}. We observe that the Random Decision Forest based model outperforms the Linear Regression based model by presenting smaller MAE and RMSE. Therefore, we employ the Random Decision Forest based model to estimate current city exposure probability, denoted as \textit{RDF Exposure Model}.

\begin{table}
\centering
\scriptsize
\begin{tabular}{c||c|c}
  \hline
 & \bfseries Random Decision Forest & \bfseries Linear Regression \\
  \hline
  \bfseries MAE & $0.027$ & $0.061$  \\
  \bfseries RMSE & $0.077$ & $0.146$ \\
  \hline
\end{tabular}
\caption{Performance Comparison of Exposure Models}
\label{table:EM_comp}
\end{table}

\begin{table}
\centering
\scriptsize
\begin{tabular}{c||c|c|c|c|c}
  \hline
  & \bfseries RDF Exposure & \bfseries No Error & \bfseries No User & \bfseries No \% Friends & \bfseries No Cluster \\
  & \bfseries Model & \bfseries Distance & \bfseries Category & \bfseries with Attributes & \bfseries Confidence\\
  \hline
  \bfseries MAE &  $0.027$ & $0.052$ & $0.065$ & $0.045$ & $0.082$ \\
  \bfseries RMSE & $0.077$  & $0.106$ & $0.131$ & $0.117$  & $0.166$  \\
  \hline
\end{tabular}
\caption{Feature Verification of \textit{RDF Exposure Model}}
\label{table:EM_features}
\end{table}

Furthermore, `Leave-one-feature-out' approach is exploited to verify the effectiveness of the features. We use Random Decision Forest approach to train exposure models by taking out any one of the four features, namely \textit{No Error Distance}, \textit{No User Category}, \textit{No \% Friends with Attributes} and \textit{No Cluster Confidence}. Table~\ref{table:EM_features} compares these `Leave-one-feature-out' models to the \textit{RDF Exposure Model}. We observe that the \textit{RDF Exposure Model} presents the best performance with the smallest MAE and RMSE. The performance degradations when removing any one of the features just verify that all the four studied features contribute to the model. \textit{Cluster Confidence} is observed as the most sensitive feature for the model, because the performance of the \textit{RDF Exposure Model} drops most significantly when \textit{Cluster Confidence} is taken out.

\subsubsection{Current City Exposure Estimator}
\begin{figure}[!htp] \centering
\includegraphics[width=8cm, height=3.8cm]{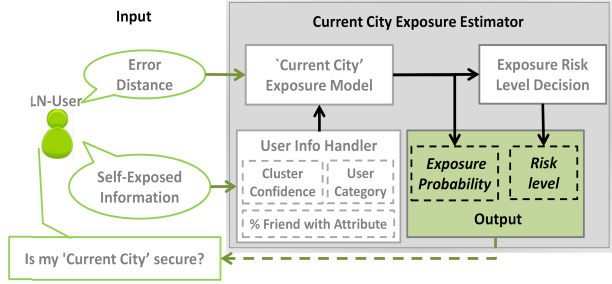}
\caption{Framework of Current City Exposure Estimator.}
\label{fig:Exposure_Reminder}
\end{figure}

By exploiting the proposed current city exposure model, we construct
an exposure estimator to forecast the exposure risk of a user's
private current city. Figure \ref{fig:Exposure_Reminder} illustrates
the framework of the current city exposure estimator. The exposure
estimator contains three main function modules: user information
handler, current city exposure model and exposure risk level
decision. The inputs of the exposure estimator include a user's
self-exposed information and a pre-established \textit{Error
Distance}. Given a user's self-exposure information, the user
information handler determines \textit{User Category}, and computes
\textit{Cluster Confidence} and \textit{\% Friends with Attributes}. Based on the pre-established
\textit{Error Distance}, the obtained \textit{User Category},
\textit{Cluster Confidence}, and \textit{\% Friends with Attributes}, the exposure model calculates the
current city exposure probability for the user. The exposure risk module
determines a risk level according to the exposure probability.
Finally, the exposure estimator outputs two risk measurements of
current city: \textit{Exposure Probability} and \textit{Risk Level}.

\subsection{Case Studies: Exposure Estimator and Privacy Protection}
\begin{table*}[!htp]
\centering
\scriptsize
\begin{tabular}{c||c|c|c|c||c|c}
  \hline
  \multirow{2}*{\bfseries User} & \bfseries User & \bfseries Cluster & \bfseries Error & \bfseries \% Friends with & \bfseries Exposure &\bfseries Risk\\
  & \bfseries Category & \bfseries Confidence& \bfseries Distance& \bfseries Attribute & \bfseries Probability & \bfseries Level\\
  \hline
  \bfseries $U1$ & `HT+WE+F' & $0.69$ &$100 km$ & $0.9\%$ & 0.967 & Level 5\\
  \bfseries $U1$ & `HT+WE+F' & $0.69$ & $20 km$ & $0.9\%$ & 0.883 & Level 4\\
  \bfseries $U2$ & `F' & $0.208$ & $100 km$ & $11.2\%$ & 0.564 & Level 3\\
  \bfseries $U3$ & `F' & $0.208$ & $100 km$ & $0.2\%$ & 0.374 & Level 2\\
  \bfseries $U4$ & `WE+F' & $0.281$ & $100 km$ & $2.1\%$ & 0.407 & Level 2\\
  \bfseries $U5$ & `WE+F' & $0.57$ & $100 km$ & $2.1\%$ & 0.797 & Level 4\\
  \bfseries $U6$ & `HT+F' & $0.332$ & $20 km$ & $20.1\%$ & 0.276 & Level 2\\
  \bfseries $U7$ & `HT+WE' & $0.73$ & $100 km$ & $0\%$ & 0.903 & Level 5\\
  \bfseries $U8$ & `HT' & $0.169$ & $20 km$ & $0\%$ & 0.059 & Level 1\\
  \bfseries $U9$ & `WE' & $0.404$ & $20 km$ & $0\%$ & 0.834 & Level 4\\
  \bfseries $U10$ & `F' & $0.891$ & $20 km$ & $17.2\%$ & 0.823 & Level 4\\
  \hline
\end{tabular}
\caption{Exposure Estimator Cases Study}
\label{table:User_Case}
\end{table*}

Any LN-users who reveal their self-exposed information and pre-define an \textit{Error Distance} can use the proposed current city exposure estimator to assess their \textit{Exposure Probability} and \textit{Risk Level}. To better understand the use of exposure estimator, we illustrate several use cases in Table \ref{table:User_Case}. In this study, we observe that some of the
LN-users are not really safe to hide their current city if they
leave some other information visible. For instance, considering
$U9$, even though only `WE' is published, his current city is almost leaked
with an extremely high \textit{Exposure Probability} of $0.834$
within an \textit{Error Distance} of $20$ km. In addition, for users
in the same \textit{User Category}, the one with a higher
\textit{Cluster Confidence} is more likely to divulge his current
city. Looking at $U4$ and $U5$ who are both in `WE+F' category, the
current city of $U5$ who exhibits a higher \textit{Cluster Confidence} is more dangerous to be inferred, compared to $U4$'s current city.

In addition, the exposure estimator can offer some countermeasures on
privacy configuration against information leakage. Assume users
hide some part of their exposed information, the exposure estimator
estimates and reports the corresponding \textit{Exposure
Probability} and \textit{Exposure Risk Level}. Then users can decide on
a new privacy configuration accordingly. We take $U1$ as an example
and list some possible exposure risks assuming that he adjusts his
privacy configuration. The results shown in Table
\ref{table:Exa_Guide} reveal that the exposure risk could be significantly decreased
if $U1$ hides his `HT+WE', `WE+F' or `WE'. The results also point
out that merely hiding `F' or `HT' cannot protect $U1$'s current city
privacy.

\begin{table}
\centering
\scriptsize
\begin{tabular}{c||c|c|c|c|c|c}
\hline

  \multicolumn{1}{c||}{\multirow{2}{*}{\bfseries $U1$}} & \multicolumn{1}{c|}{\bfseries Current status} & \multicolumn{5}{c}{\bfseries Hide}  \\
  \cline{3-7}
  \multicolumn{1}{c||}{}  & \multicolumn{1}{c|}{\bfseries `HT+WE+F'} & \multicolumn{1}{c|}{\bfseries `WE'}  & \multicolumn{1}{c|}{\bfseries `F'} & \multicolumn{1}{c|}{\bfseries `HT'} & \multicolumn{1}{c|}{\bfseries `WE+F'} & \multicolumn{1}{c}{\bfseries `HT+WE'}\\  %

  \hline
  \bfseries Exposure  & \multirow{2}*{$0.967$} & \multirow{2}*{$0.503$} & \multirow{2}*{$0.944$} & \multirow{2}*{$0.936$} & \multirow{2}*{$0.456$} & \multirow{2}*{$0.073$} \\
  \bfseries Probability &  &  &  &  &  & \\
  \hline
  \bfseries Risk Level & Level 5 & Level 3 & Level 5 & Level 5 & Level 2 & Level 1\\
  \hline
\end{tabular}
\caption{Exposure Guidelines for $U1$: the exposure risks if he
adjusts some privacy configurations with an \textit{Error Distance}
of $100km$}
\label{table:Exa_Guide}
\end{table}

Finally, according to the studies on current city exposure
risk, we summarize the following general suggestions:
\begin{itemize}
  \item As all the location indications may expose the hidden current city, close all of location sensitive information including `WE', `F' and `HT' so as to achieve a high current city security.
  \item Hide the most sensitive exposed information (e.g., `WE') if users want to publicly share some personal information (e.g., `F'), since the most
  sensitive information can independently lead to a quite high \textit{Exposure Probability}. For example, `WE' alone can lead to an \textit{Exposure Probability} higher than $80\%$.
  \item According to the centrality principle which refers to the \textit{Cluster Confidence}, hide `F' if most friends indicate the same place where the user
  lives. For instance, $U10$ in Table \ref{table:User_Case} is
  necessarily advised to hide his `F'.
\end{itemize}

\section{Discussion and Future Work}
\label{sec:Discussion}
In this section, we discuss some issues which are not addressed in this work due to space limitations, and point out some future potential research directions.

\subsection{Extensibility of the Current City Prediction Approach}
\label{subsec:extensibility_of_prediction_approach}
Due to the data set limitation, we only use three features (i.e., `Hometown', `Work and Education' and `Friend') to evaluate our proposed current city prediction approach. However, our prediction approach can be extended to consider other location sensitive attributes. For instance, for the location sensitive pages that a user follows (e.g., the page of a favorite local restaurant) or the location sensitive posts that a user published (e.g., geo-tagged posts), we can regard one page or one post as a LA-Friend and refer to \textit{LA-FLI} model to explore the location indications.

\subsection{Adaptability of the Exposure Estimation Approach}
In addition, our exposure estimation approach can easily adapt to other current city prediction approaches by the two-step solution: (1) \textit{feature extraction} (Sec.~\ref{sec:expo_inspection}) and (2) \textit{exposure model training} (Sec.~\ref{expo_extimation}). In particular, we can first extract similar features for other city prediction approaches as the inspected features in Sec.~\ref{sec:expo_inspection}. 
Take \textit{Cluster Confidence} as an example. For the cluster-based city prediction approaches like ours, \textit{Cluster Confidence} can be extracted in the same way, i.e., the largest \textit{cluster} prediction probability (Eq.\ref{eq:cluster_confidence}). For the other city prediction approaches without a clustering step~\cite{find_me}\cite{UID}, following the essence of \textit{Cluster Confidence}, a similar feature, \textit{Prediction Confidence}, can be computed as the largest \textit{city} prediction probability. Likewise, we can also obtain the other features presented in our exposure model for many other city prediction approaches, while we do not discuss them further for brevity. Once the features are derived, in the second step, we can directly apply the regression methods used in Sec.~\ref{expo_extimation} to train the exposure models for other prediction approaches.


\subsection{Generalizability of the Exposure Estimator}
Taking `current city' as a representative attribute to study the information exposure issue, this work gives further insights on how to assess the exposure risk of other privacy-sensitive attributes (e.g., age). Denoting the privacy-sensitive attribute as PSA, the process to assess its exposure risk can be generalized into three steps: $1)$ Explore PSA-sensitive attributes and construct a PSA prediction model; $2)$ Inspect the prediction results to extract features and train a PSA exposure model; $3)$ Based on the exposure model, implement an exposure estimator to notify users of the exposure risk and provide suggestions to lower the risk if necessary.

Moreover, our future work will consider integrating multiple exposure models into the exposure estimator, so as to construct an exposure estimation system that can provide reliable and multi-functional exposure risk estimations.

\section{Conclusion}
\label{sec:conclusion} This paper starts with two open questions
regarding the security of users' hidden privacy-sensitive
attributes. To answer these questions, we first propose a novel
current city prediction approach to infer users' current city by
leveraging users' self-exposed information including location
sensitive attributes and friends list. We validate the new
prediction approach on a Facebook data set containing $371,913$ users,
and the results reveal that the users' hidden current city may be
dangerous to be predicted. Then we apply the proposed prediction
approach to predict users' current city and model the exposure
probability by considering four measurable characteristics --- \textit{Cluster Confidence}, \textit{Error Distance}, \textit{User Category} and \textit{Percentage of Friends with Attributes}. Based on
the exposure model, we propose a current city exposure estimator to measure
the exposure probability and risk level of users' hidden current city
according to their self-exposed information. The exposure estimator
can also help users to adjust their privacy configuration to satisfy
their privacy requirements. While this work studies the
potential risk of users' privacy-sensitive attributes with a
representative attribute of current city in Facebook, the proposed
idea and approach could be extended to other attributes and
utilized by other OSNs.

\section*{Acknowledgment}
We would like to acknowledge Dr. Rebecca Copeland for the careful proof-reading. This work has been funded by the China Scholarship Council.

\nocite{*}
\bibliographystyle{elsarticle-harv}
\bibliography{relatedwork}








\end{document}